\input epsf
\newfam\scrfam
\batchmode\font\tenscr=rsfs10 \errorstopmode
\ifx\tenscr\nullfont
        \message{rsfs script font not available. Replacing with calligraphic.}
        \def\scr{\cal}
\else   
        \font\sevenscr=rsfs7
        \font\fivescr=rsfs5
        \skewchar\tenscr='177 \skewchar\sevenscr='177 \skewchar\fivescr='177
        \textfont\scrfam=\tenscr \scriptfont\scrfam=\sevenscr
        \scriptscriptfont\scrfam=\fivescr
        \def\scr{\fam\scrfam}
        
\fi
\catcode`\@=11
\newfam\frakfam
\batchmode\font\tenfrak=eufm10 \errorstopmode
\ifx\tenfrak\nullfont
        \message{eufm font not available. Replacing with italic.}
        \def\frak{\it}
\else
    
    \font\sevenfrak=eufm7 \font\fivefrak=eufm5
    \textfont\frakfam=\tenfrak
    \scriptfont\frakfam=\sevenfrak \scriptscriptfont\frakfam=\fivefrak
    \def\frak{\fam\frakfam}
\fi
\catcode`\@=\active
\newfam\msbfam
\batchmode\font\twelvemsb=msbm10 scaled\magstep1 \errorstopmode
\ifx\twelvemsb\nullfont\def\Bbb{\bf}

    \message{Blackboard bold not available. Replacing with boldface.}
\else   \catcode`\@=11
        \font\tenmsb=msbm10 \font\sevenmsb=msbm7 \font\fivemsb=msbm5
        \textfont\msbfam=\tenmsb
        \scriptfont\msbfam=\sevenmsb \scriptscriptfont\msbfam=\fivemsb
        \def\Bbb{\relax\expandafter\Bbb@}
        \def\Bbb@#1{{\Bbb@@{#1}}}
        \def\Bbb@@#1{\fam\msbfam\relax#1}
        \catcode`\@=\active

\fi
        \font\eightrm=cmr8              \def\xrm{\eightrm}
        \font\eightbf=cmbx8             \def\xbf{\eightbf}
        \font\eightit=cmti10 at 8pt     \def\xit{\eightit}
        \font\sixit=cmti6               \def\xxit{\sixit}
        \font\sixrm=cmr6                \def\xxrm{\sixrm}
                     
        \font\eightcp=cmcsc8
        \font\eighti=cmmi8              \def\xold{\eighti}
        \font\eightib=cmmib8             \def\xbold{\eightib}
        \font\teni=cmmi10               \def\old{\teni}
        \font\tencp=cmcsc10

        \font\twelvecp=cmcsc10 scaled\magstep1

        \font\eightmath=cmmi8
        \font\sixmath=cmmi6

\batchmode\font\tenhelvbold=phvb at10pt \errorstopmode
\ifx\tenhelvbold\nullfont
        \message{phvb font not available. Replacing with cmr.}
    \font\tenhelvbold=cmb10   
    \font\twelvehelvbold=cmb12
    
    \font\sixteenhelvbold=cmb16
  \else
    \font\tenhelvbold=phvb at10pt   
    \font\twelvehelvbold=phvb at12pt
     at14pt
    \font\sixteenhelvbold=phvb at16pt
\fi

\def\noblackbox{\overfullrule=0pt}
\noblackbox

\newtoks\headtext
\headline={\ifnum\pageno=1\hfill\else
    \ifodd\pageno{\eightcp\the\headtext}{ }\dotfill{ }{\old\folio}
    \else{\old\folio}{ }\dotfill{ }{\eightcp\the\headtext}\fi
    \fi}
\def\makeheadline{\vbox to 0pt{\vss\noindent\the\headline\break
\hbox to\hsize{\hfill}}
        \vskip2\baselineskip}
\newcount\infootnote
\infootnote=0
\def\foot#1#2{\infootnote=1
\footnote{${}^{#1}$}{\vtop{\baselineskip=.75\baselineskip
\advance\hsize by 
-\parindent\noindent{\xrm #2\hfill\vskip\parskip}}}\infootnote=0$\,$}
\newcount\refcount
\refcount=1
\newwrite\refwrite
\def\oldsize{\ifnum\infootnote=1\xold\else\old\fi}
\def\ref#1#2{
    \def#1{{{\oldsize\the\refcount}}\ifnum\the\refcount=1\immediate\openout\refwrite=\jobname.refs\fi\immediate\write\refwrite{\item{[{\xold\the\refcount}]}
    #2\hfill\par\vskip-2pt}\xdef#1{{\noexpand\oldsize\the\refcount}}\global\advance\refcount by 1}
    }
\def\refout{\catcode`\@=11
        \xrm\immediate\closeout\refwrite
        \vskip2\baselineskip
        {\noindent\twelvecp References}\hfill
        \par\nobreak\vskip\baselineskip
        \baselineskip=.75\baselineskip
        \input\jobname.refs
        \baselineskip=4\baselineskip \divide\baselineskip by 3
        \catcode`\@=\active\rm}

\def\hepth#1{\href{http://arxiv.org/abs/hep-th/#1}{arXiv:hep-th/{\xold#1}}}
\def\arxiv#1#2{\href{http://arxiv.org/abs/#1.#2}{arXiv:{\xold#1}.{\xold#2}}}
\def\jhep#1#2#3#4{\href{http://jhep.sissa.it/stdsearch?paper=#2\%28#3\%29#4}{J. High Energy Phys. {\xbold #1#2} ({\xold#3}) {\xold#4}}}

\def\CMP#1#2#3{Commun. Math. Phys. {\xbold#1} ({\xold#2}) {\xold#3}}
\def\CQG#1#2#3{Class. Quantum Grav. {\xbold#1} ({\xold#2}) {\xold#3}}
\def\GRG#1#2#3{Gen. Rel. Grav. {\xbold#1} ({\xold#2}) {\xold#3}}

\def\JHEP{\jhep}

\def\NPB#1#2#3{Nucl. Phys. {\xbf B}{\xbold#1} ({\xold#2}) {\xold#3}}

\def\PLB#1#2#3{Phys. Lett. {\xbf B}{\xbold#1} ({\xold#2}) {\xold#3}}
\def\PR#1#2#3{Phys. Rept. {\xbold#1} ({\xold#2}) {\xold#3}}
\def\PRD#1#2#3{Phys. Rev. {\xbf D}{\xbold#1} ({\xold#2}) {\xold#3}}
\def\PRL#1#2#3{Phys. Rev. Lett. {\xbold#1} ({\xold#2}) {\xold#3}}

\newcount\sectioncount
\sectioncount=0
\def\section#1#2{\global\eqcount=0
    \global\subsectioncount=0
        \global\advance\sectioncount by 1
    \ifnum\sectioncount>1
            \vskip2\baselineskip
    \fi
    \noindent
        \line{\twelvecp\the\sectioncount. #2\hfill}
        \par\nobreak\vskip.8\baselineskip\noindent
        \xdef#1{{\old\the\sectioncount}}}
\newcount\subsectioncount
\def\subsection#1#2{\global\advance\subsectioncount by 1
    \par\nobreak\vskip.8\baselineskip\noindent
    \line{\tencp\the\sectioncount.\the\subsectioncount. #2\hfill}
    \vskip.5\baselineskip\noindent
    \xdef#1{{\old\the\sectioncount}.{\old\the\subsectioncount}}}
\newcount\appendixcount
\appendixcount=0
\def\appendix#1{\global\eqcount=0
        \global\advance\appendixcount by 1
        \vskip2\baselineskip\noindent
        \ifnum\the\appendixcount=1
        \hbox{\twelvecp Appendix A: #1\hfill}
        \par\nobreak\vskip\baselineskip\noindent\fi
    \ifnum\the\appendixcount=2
        \hbox{\twelvecp Appendix B: #1\hfill}
        \par\nobreak\vskip\baselineskip\noindent\fi
    \ifnum\the\appendixcount=3
        \hbox{\twelvecp Appendix C: #1\hfill}
        \par\nobreak\vskip\baselineskip\noindent\fi}
\def\acknowledgements{\vskip2\baselineskip\noindent
        \underbar{\it Acknowledgements:}\ }
\newcount\eqcount
\eqcount=0
\def\Eqn#1{\global\advance\eqcount by 1
\ifnum\the\sectioncount=0
    \xdef#1{{\old\the\eqcount}}
    \eqno({\oldstyle\the\eqcount})
\else
        \ifnum\the\appendixcount=0
            \xdef#1{{\old\the\sectioncount}.{\old\the\eqcount}}
                \eqno({\oldstyle\the\sectioncount}.{\oldstyle\the\eqcount})\fi
        \ifnum\the\appendixcount=1
            \xdef#1{{\oldstyle A}.{\old\the\eqcount}}
                \eqno({\oldstyle A}.{\oldstyle\the\eqcount})\fi
        \ifnum\the\appendixcount=2
            \xdef#1{{\oldstyle B}.{\old\the\eqcount}}
                \eqno({\oldstyle B}.{\oldstyle\the\eqcount})\fi
        \ifnum\the\appendixcount=3
            \xdef#1{{\oldstyle C}.{\old\the\eqcount}}
                \eqno({\oldstyle C}.{\oldstyle\the\eqcount})\fi
\fi}
\def\eqn{\global\advance\eqcount by 1
\ifnum\the\sectioncount=0
    \eqno({\oldstyle\the\eqcount})
\else
        \ifnum\the\appendixcount=0
                \eqno({\oldstyle\the\sectioncount}.{\oldstyle\the\eqcount})\fi
        \ifnum\the\appendixcount=1
                \eqno({\oldstyle A}.{\oldstyle\the\eqcount})\fi
        \ifnum\the\appendixcount=2
                \eqno({\oldstyle B}.{\oldstyle\the\eqcount})\fi
        \ifnum\the\appendixcount=3
                \eqno({\oldstyle C}.{\oldstyle\the\eqcount})\fi
\fi}
\def\multi{\global\advance\eqcount by 1}
\def\multieq#1#2{
    \ifnum\the\sectioncount=0
        \eqno({\oldstyle\the\eqcount})
         \xdef#1{{\old\the\eqcount#2}}
    \else
        \xdef#1{{\old\the\sectioncount}.{\old\the\eqcount}#2}
        \eqno{({\oldstyle\the\sectioncount}.{\oldstyle\the\eqcount}#2)}
    \fi}

\newtoks\url
\def\Href#1#2{\catcode`\#=12\url={#1}\catcode`\#=\active#2}
\def\href#1#2{{#2}}

\parskip=3.5pt plus .3pt minus .3pt
\baselineskip=14pt plus .1pt minus .05pt
\lineskip=.5pt plus .05pt minus .05pt
\lineskiplimit=.5pt
\abovedisplayskip=18pt plus 4pt minus 2pt
\belowdisplayskip=\abovedisplayskip
\hsize=14cm
\vsize=19cm
\hoffset=1.5cm
\voffset=1.8cm
\frenchspacing
\footline={}
\raggedbottom

\def\ss{\scriptstyle}
\def\sss{\scriptscriptstyle}
\def\*{\partial}
\def\punkt{\,\,.}
\def\komma{\,\,,}

\def\={\!=\!}
\def\small#1{{\hbox{$#1$}}}

\def\fraction#1{\small{1\over#1}}
\def\fr{\fraction}
\def\Fraction#1#2{\small{#1\over#2}}
\def\Fr{\Fraction}
\def\tr{\hbox{\rm tr}}
\def\eg{{\tenit e.g.}}

\def\ie{{\tenit i.e.}}

\def\nlni{\hfill\break}

\def\a{\alpha}

\def\d{\delta}
\def\e{\varepsilon}
\def\g{\gamma}

\def\m{\mu}

\def\gg{{\frak g}}
\def\kk{{\frak k}}

\def\G{\Gamma}

\def\Z{{\Bbb Z}}

\def\id{1\hskip-3.5pt 1}


\def\ee{{\frak e}}

\def\rdiag#1{\hbox{\epsfxsize=12pt\lower3pt\hbox{\epsffile{#1.eps}}}}
\def\rrrdiag#1{\hbox{\epsfxsize=12pt\lower3pt\hbox{\epsffile{r3-#1.eps}}}}

\def\eeR{\e\e\hat R^4}
\def\ttR{t_8t_8\hat R^4}

\def\i{\hbox{\it i}}
\def\ii{\hbox{\it ii}}
\def\iii{\hbox{\it iii}}
\def\iv{\hbox{\it iv}}
\def\v{\hbox{\it v}}
\def\vi{\hbox{\it vi}}
\def\vii{\hbox{\it vii}}
\def\viii{\hbox{\it viii}}
\def\ix{\hbox{\it ix}}
\def\x{\hbox{\it x}}

\def\P{{\scr P}}
\def\II{I\hskip-.5pt I}

\def\ms{{\mathstrut}}


\ref\CederwallPalmkvist{M. Cederwall and J. Palmkvist, {\xit ``The
octic E${}_{\sss8}$ invariant''}, \hepth{0702024}, J.
Math. Phys. in press.}

\ref\Fulling{S.A. Fulling, R.C. King, B.G. Wybourne and C.J.
Cummins, {\xit "Normal forms for tensor polynomials: I. The Riemann
tensor"}, \CQG {9}{1992}{1151}.}

\ref\GreenVanhoveI{M.B. Green and P. Vanhove, {\xit ``D-instantons,
strings and M-theory''}, \PLB{408}{1997}{122} \nlni
[\hepth{9704145}].}

\ref\GreenVanhoveII{M.B. Green and P. Vanhove, {\xit ``Duality and
higher derivative terms in M theory''}, \jhep{06}{01}{2006}{093}
[\hepth{0510027}].}

\ref\GreenGutVan{M.B. Green, M. Gutperle and P. Vanhove, {\xit ``One
loop in eleven dimensions''}, \PLB{409}{1997}{177} [\hepth{9706175}].}

\ref\GreenKwonVan{M.B. Green, H.-h. Kwon and P. Vanhove, {\xit ``Two
loops in eleven dimensions''}, \PRD{61}{2000}{104010} [\hepth{9910055}].}

\ref\CremmerPopeI{E. Cremmer, B. Julia, H. L\"u and C.N. Pope,
{\xit ``Dualisation of dualities. I.''}, \NPB{523}{1998}{73} [\hepth{9710119}].}

\ref\HyakutakeOgushi{Y. Hyakutake and S. Ogushi,
{\xit ``R\raise3pt\hbox{\xxit4} corrections to eleven dimensional supergravity
via supersymmetry''}, \PRD{74}{2006}{025022} [\hepth{0508204}];
{\xit ``Higher derivative corrections to eleven dimensional
  supergravity via local supersymmetry''},
\jhep{06}{02}{2006}{068} [\hepth{0601092}].}

\ref\HigherDer{T. Damour and H. Nicolai, {\xit ``Higher order M
theory corrections and the Kac--Moody algebra E\lower2pt\hbox{\xxit10}''},
\CQG{22}{2005}{2849} [\hepth{0504153}];  T. Damour, A. Hanany, M.
Henneaux, A. Kleinschmidt and H. Nicolai, {\xit ``Curvature
corrections and Kac--Moody compatibility conditions''},
\GRG{38}{2006}{1507} [\hepth{0604143}].}

\ref\LambertWestI{N. Lambert and P. West, {\xit ``Enhanced coset
symmetries and higher derivative corrections''},
\PRD{74}{2006}{065002} [\hepth{0603255}].}

\ref\LambertWestII{N. Lambert and P. West, {\xit ``Duality groups,
automorphic forms and higher derivative corrections''},
\PRD{75}{2007}{066002} [\hepth{0611318}].}

\ref\Meissner{K.A. Meissner, {\xit ``Symmetries of higher-order
string gravity actions''}, \PLB{392}{1997}{298}
\nlni[\hepth{9610131}].}

\ref\MizoguchiSchroder{S. Mizoguchi and G. Schr\"oder, {\xit ``On
discrete U-duality in M-theory''}, \CQG{17}{2000}{835}
[\hepth{9909150}].}

\ref\NilssonTollstenIII{B.E.W. Nilsson and A. Tollst\'en, {\xit
"Supersymmetrization of 
{\eightmath\char16}(3)R\raise3pt\hbox{\xxit4}\hskip-4pt
\lower2pt\hbox{\sixmath\char22\char23\char26\char27} in
superstring theories"}, \PLB{181}{1986}{63}.}

\ref\LuPope{H. L\"u and C.N. Pope, {\xit ``p-brane solitons in
maximal supergravities''}, \NPB{465}{1996}{127} [\hepth{9512012}].}

\ref\RussoTseytlin{J.G. Russo and A.A. Tseytlin, {\xit "One-loop
four-graviton amplitude in eleven-dimensional supergravity"},
\NPB{508}{1997}{261} [\hepth{9707134}].}

\ref\SixteenFermions{M.B. Green, M. Gutperle and H. Kwon, {\xit
``Sixteen-fermion and related terms in M-theory on
T\raise3pt\hbox{\xxit2}''}, \PLB{421}{1998}{149}
[\hepth{9710151}].}

\ref\SheikhJabbari{Q. Exirifard and M.M. Sheikh-Jabbari, {\xit
``Lovelock gravity at the crossroads of Palatini and metric
formulations''}, \arxiv{0705}{1879}.}

\ref\TsimpisPolicastro{G. Policastro and D. Tsimpis,
{\xit ``R\raise3pt\hbox{\xxit 4}, purified''},
\CQG{23}{2006}{4753} [\hepth{0603165}].}

\ref\UdualityMembranes{V. Bengtsson, M. Cederwall, H. Larsson and
B.E.W. Nilsson, {\xit ``U-duality covariant
membranes''}, \jhep{05}{02}{2005}{020} [\hepth{0406223}].}

\ref\PetersVanhoveWesterberg{K. Peeters, P. Vanhove and A.
Westerberg, {\xit ``Supersymmetric higher-derivative actions in ten
and eleven dimensions, the associated superalgebras and their
formulation in superspace''}, \CQG{18}{2001}{843}
[\hepth{0010167}].}

\ref\ObersPioline{N.A. Obers and B. Pioline, {\xit ``Eisenstein series and
string thresholds''}, \CMP{209}{2000}{275} [\hepth{9903113}].}

\ref\ObersPiolineU{N.A. Obers and B. Pioline, {\xit ``U-duality and M-theory''},
\PR{318}{1999}{113}
\nlni [\hepth{9809039}].}

\ref\DamourHenneauxNicolai{T. Damour, M. Henneaux and H. Nicolai,
{\xit ``E(10) and a 'small tension expansion' of M theory''},
\PRL{89}{2002}{221601} [\hepth{0207267}].}

\ref\DamourNicolai{T. Damour and H. Nicolai,
{\xit ``Symmetries, singularities and the de-emergence of space''},
\arxiv{0705}{2643}.}

\ref\WestEeleven{P. West, {\xit ``E(11) and M theory''},
\CQG{18}{2001}{4443} [\hepth{0104081}].}

\ref\EHTP{F. Englert, L. Houart, A. Taormina and P. West,
{\xit ``The symmetry of M theories''},
\jhep{03}{09}{2003}{020} [\hepth{0304206}].}

\ref\CGNN{M. Cederwall, U. Gran, M. Nielsen and B.E.W. Nilsson,
{\xit ``Manifestly supersymmetric M-theory''},
\JHEP{00}{10}{2000}{041} [\hepth{0007035}]; {\xit ``Generalised
11-dimensional supergravity''}, in proceedings of ``Quantization,
Gauge Theory and Strings'', Moscow 2000, eds. A. Semikhatov, M.
Vasiliev and V. Zaikin [\hepth{0010042}].}

\ref\CGNT{M. Cederwall, U. Gran, B.E.W. Nilsson and D. Tsimpis,
{\xit ``Supersymmetric corrections to eleven-dimen\-sional supergravity''},
\jhep{05}{05}{2005}{052} [\hepth{0409107}].}

\ref\HoweRfour{P.S. Howe, {\xit ``R\raise3pt\hbox{\xxit 4} terms
in supergravity and M-theory''}, \hepth{0408177}.}

\ref\GreenSethi{M.B.~Green and S.~Sethi,
{\xit ``Supersymmetry constraints on type IIB supergravity''},
\PRD{59}{1999}{046006} [\hepth{9808061}].}

\ref\KazhdanPiolineWaldron{D. Kazhdan, B. Pioline and A. Waldron,
{\xit ``Minimal representations, spherical vectors,
and exceptional theta series. I.''}, \CMP{226}{2002}1 [\hepth{0107222}].}

\headtext={Bao, Cederwall, Nilsson: ``Aspects of
Higher Curvature Terms and U-Duality''}

\line{
\epsfysize=15mm
\epsffile{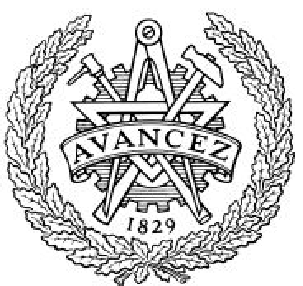}
\hfill}
\vskip-12mm
\line{\hfill G\"oteborg preprint}
\line{\hfill June, {\old2007}}
\line{\hrulefill}

\vfill
\vskip.5cm

\centerline{\sixteenhelvbold
Aspects of Higher Curvature Terms}

\vskip.3cm
\centerline{\sixteenhelvbold and U-Duality}

\vfill

\centerline{\twelvehelvbold
Ling Bao, Martin Cederwall and Bengt EW Nilsson}

\vfill

\centerline{\it Fundamental Physics}
\centerline{\it Chalmers University of Technology}
\centerline{\it SE 412 96 G\"oteborg, Sweden}

\vfill

{\narrower\noindent \underbar{Abstract:} We discuss various aspects
of dimensional reduction of gravity with the Einstein--Hilbert action
supplemented by a lowest order
deformation formed as the Riemann tensor raised to powers
two, three or four. In the case of $R^2$ we give an explicit 
expression, 
and discuss the
possibility of extended coset symmetries, especially $SL(n+1,\Z)$
for reduction on an $n$-torus to three dimensions. Then we start an
investigation of the dimensional reduction of $R^3$ and $R^4$ by
calculating some terms relevant for the coset formulation, aiming in
particular towards $E_{8(8)}/(Spin(16)/\Z_2)$ in three dimensions and
an investigation of the derivative structure. We
emphasise some issues concerning the need for the introduction of
non-scalar automorphic forms in order to realise certain expected
enhanced symmetries.
\smallskip}
\vfill

\font\xxtt=cmtt6

\vtop{\baselineskip=.6\baselineskip\xxtt
\line{\hrulefill}
\catcode`\@=11
\line{email: ling.bao@chalmers.se, martin.cederwall@chalmers.se,
tfebn@fy.chalmers.se\hfill}
\catcode`\@=\active
}

\eject

\section\Intro{Introduction and summary}M-theory, when compactified on
an $n$-torus, is conjectured to have a global U-duality symmetry $E_{n(n)}$ in
the low-energy limit described by maximal
supergravity in $d=11-n$ dimensions. It is known from string theory that
this continuous symmetry is broken in the quantum theory to a discrete
version $E_{n(n)}(\Z)$. The massless scalars in the compactified theory
belong to the coset $E_{n(n)}(\Z)\backslash E_{n(n)}/K(E_{n(n)})$, where
$K(E_{n(n)})$ is the (locally implemented) maximal compact subgroup of
the split form $E_{n(n)}$.
When $d\leq3$, no local massless bosonic degrees of freedom remain except
scalar ones.
It has been proposed that it may even be possible
to define M-theory itself as a theory on the coset obtained when going
to $d=1$ ($E_{10}$) or $d=0$ ($E_{11}$), although it is unclear whether or not
such a formulation incorporates degrees of freedom beyond supergravity.

Some aspects of these discrete symmetries are well investigated.
This concerns primarily calculations in cases with low dimension of
the torus. For $n<3$, non-perturbative string theory results are
obtained from loop calculations in $D=11$ supergravity. For $n\geq3$
one expects that there will be contributions also from membrane
instantons and for $n\geq6$ from five-brane instantons. This makes
results for higher-dimensional tori harder to obtain. On the other
hand, one may turn the argument around and ask what kind of
restrictions U-duality puts on the possible quantum corrections of
the theory. It is convenient to work in the massless sector,
obtained by dimensional reduction, and let quantum effects manifest
themselves in an effective action, which will then contain higher
orders of curvatures (and other fields), \ie, higher-derivative
terms.

Some partial results have been obtained by investigating the general
structures of higher-derivative terms to determine if they can be
made to fit into something U-duality invariant. For example it has
been shown that the Riemann tensor in $D=11$ comes only in powers
$3k+1$, where $k$ is integer. The purpose of the present paper is to
initiate a more detailed analysis aiming at actually checking the
invariance. The scope of the paper is modest; we restrict our
attention to the $D$-dimensional gravitational sector alone. Then we
set out to form higher-derivative corrections to the
Einstein--Hilbert action in the form of second, third and fourth
powers of the Riemann tensor. The full U-duality group is not
accessible with gravity only, but on compactification to $d=3$ there
still has to be an enhancement from $SL(8)$ to $SL(9)$, which is the
subgroup of $E_{8(8)}$ of which the gravitational scalars form a
coset (more generally, on reduction from $n+3$ to 3 dimensions, we
expect an enhancement from $SL(n)$ to $SL(n+1)$). Some aspects about
the general structures of the higher curvature terms at hand are
investigated, before we turn to examining chosen subsets of terms
and thereby extracting concrete information concerning the
possibility of implementing $SL(n+1)$. We draw some definite
conclusions about the necessity of introducing transforming
automorphic forms, and show that they can always be chosen to
reproduce the results in the dimensionally reduced theory. The
interpretation of the dimensionally reduced actions is not as
U-duality invariant object {\it per se}, but as properly taken large
volume limits of U-duality invariant actions involving transforming
automorphic forms. The investigation is very much a partial one, and
we point out some further directions, such as a more complete
expansion in fields, and a concrete examination of cosets and
discrete groups based on exceptional groups.

We refer to refs. [\CremmerPopeI,\ObersPiolineU] for an overview of
U-duality. Topics on $E_{10}$ and $E_{11}$ as fundamental symmetries
are dealt with in refs.
[\DamourHenneauxNicolai,\DamourNicolai,\WestEeleven,\EHTP] and
references therein. Recent developments concerning the connections
between U-duality and higher curvature terms are found in refs.
[\GreenVanhoveII,\HigherDer,\LambertWestI,\LambertWestII]. For
different approaches to higher curvature terms in supergravity and
string theory, see refs. [\GreenSethi-\hbox to 5cm{
\phantom{\PetersVanhoveWesterberg,\CGNN,\CGNT,\HoweRfour,
\TsimpisPolicastro,\GreenVanhoveII}\hfill}\hskip-5cm\HyakutakeOgushi].
The $3k+1$ restriction on powers of the Riemann tensor in
eleven-dimensional supergravity is discussed in ref. [\RussoTseytlin].

\section\Compansatz{The torus dimensional reduction procedure}Our Ansatz
for dimensional reduction on an $n$-torus to three dimensions is
given by
$$
\hat{E}^{a}=e^{-\phi}e^{a}\komma\qquad\hat{E}^{i}
=(dy^{\mu}-A^{\mu}){e_{\mu}}^{i} \punkt\eqn
$$
Here, $e^\phi$ is not an independent field, but the determinant of
the internal vielbein $e_\m{}^i$. The prefactor $e^{-\phi}$ is
chosen so that a canonically normalized Einstein--Hilbert term
results in three dimensions from the reduction of such a term in the
higher-dimensional theory. Our conventions are such that $D=d+n$
with $D$ the space-time dimension before the dimensional reduction
and $d$ the one after, with $n$ the dimension of the internal torus
on which we are performing the dimensional reduction. Flat indices
are denoted $a,b,\ldots$ in space-time and $i,j,\ldots$ on the
internal manifold which is parametrised by coordinates $y^{\mu}$.
The one-forms $A^{\mu}$ in the above Ansatz are the $n$ graviphoton
potentials while ${e_{\mu}}^{i}$ is the internal vielbein and hence
an element of $GL(n)$. One of our goals will be to see if this
global symmetry (or, strictly speaking, $SL(n,\Z)$, the mapping
class group of the internal torus) is extended to larger groups when
considering Lagrangians which consist of the Einstein--Hilbert term
plus terms containing the Riemann tensor raised to powers 2, 3 and
4. This issue has previously been investigated by the authors of
ref. [\LambertWestI] where the root and weight structure of the
scalar prefactors arising in the reduction are studied. These
prefactors are in ref. [\LambertWestI] extracted by applying some
general arguments about the properties of higher derivative terms.
In a continued work [\LambertWestII] they conclude that when weights
instead of roots occur in the scalar exponent prefactors this should
be compensated for by tensorial automorphic forms. The results
obtained here by explicitly computing some of the relevant terms in
the dimensional reduction lend further support to such a
construction. Automorphic forms of $SL(2,\Z)$ with similar
non-trivial properties have already been seen to arise in the type
\II B superstring multiplying a term containing the product of 16
dilatinos [\SixteenFermions].

From the above Ansatz one easily obtains, using the zero torsion
condition, the dimensionally reduced form of the spin connection
one-form and from it the Riemann tensor two-form. By reading off the
components of these tensors using the basis indicated by the Ansatz
above, \ie,
$\hat{e}^{a}=e^{a},\hat{e}^{i}=(dy^{\mu}-A^{\mu}){e_{\mu}}^{i}$, we
get an answer without explicit graviphoton potentials since this
basis is manifestly translation invariant on the torus
[\UdualityMembranes]. In order to examine the possibility of
symmetry enhancement in reduction to $d=3$, we need the following
expressions for the components of the Riemann tensor
$$
\eqalign{ \hat{R}_{ab}{}^{cd}&=
     e^{2\phi}\bigl[R_{ab}{}^{cd}
          +4\d_{[a\ms}^{[c\ms}D^{\ms}_{b\ms]}\phi D^{d]}\phi
          +4\d_{[a\ms}^{[c\ms}D_{b]\ms}^{\ms}\phi D^{d]\ms}\phi
          -2\d_{ab\ms}^{cd\ms}D_{e\ms}^{\ms}\phi D_{\ms}^{e\ms}\phi\bigr]\cr
     &\qquad-e^{4\phi}\bigl[\fr2(F_{ab}F^{cd})+\fr2(F_{[a}{}^cF_{b]}{}^d)\bigr]
          \komma\cr
\hat{R}_{ab}{}^{cl}&=e^{3\phi}\bigl[\fr2D^cF_{ab}{}^l+D^c\phi F_{ab}{}^l
          -D_{[a}\phi F_{b]}{}^{cl}+\d_{[a}^cD^d\phi F_{b]d}{}^l\cr
     &\qquad+\fr2(F_{ab}P^c)^l+(F_{[a}{}^cP_{b]})^l\bigr]\komma\cr
\hat{R}_{ab}{}^{kl}&=-2e^{2\phi}(P_{[a}P_{b]})^{kl}
           -\fr2e^{4\phi}F_a{}^{c[k}F_{bc}{}^{l]}\komma\cr
\hat{R}_{aj}{}^{cl}&=+\fr4e^{4\phi}F^{ce}{}_jF_{ae}{}^l
           -e^{2\phi}\bigl[D_aP^c+D_a\phi P^c+D^c\phi P_a
           -\d_{a\ms}^{c\ms}D_{e\ms}^{\ms}\phi P^e+P_aP^c\bigr]_j{}^l\komma\cr
\hat{R}_{aj}{}^{kl}&=-e^{3\phi}F_a{}^{e[k}P_{ej}{}^{l]}\komma\cr
\hat{R}_{ij}{}^{kl}&=-2e^{2\phi}(P_e){}_{[i}{}^k(P^e){}_{j]}{}^l
 \komma\cr
 }\Eqn\RiemannTensor
$$
where $F^{i}_{ab}:=F^{\mu}_{ab}e^\ms_{\mu\ms}{}^{i}$, with 
$F^{\mu}_{mn}=2\partial_{[m}A_{n]}{}^\m$,
are the graviphoton field strengths. We use the notation
$(AB)=A^iB^i$ for the scalar product of $SO(n)$ vectors.
The covariant derivative is
$D_{m}=\partial_{m}+\omega_{m}+Q_{m}$. We have also defined $P$ and
$Q$ as the symmetric and antisymmetric parts of the Maurer--Cartan
one-form constructed from the internal vielbein ${e_{\mu}}^{i}$
(remember that they form the Maurer--Cartan form of $GL(n)$, so that
$\tr P=d\phi$). $Q$ belongs to the $so(n)$ subalgebra and $P$ spans
the tangent directions of the corresponding coset $GL(n)/SO(n)$. As
a direct consequence of their definition $P$ and $Q$ satisfy
$$
DP:=dP+PQ+QP=0\komma\qquad F_{Q}:=dQ+Q^2=-P^2\punkt\eqn
$$
We also have that the graviphotons satisfy the Bianchi identity
$DF-F\wedge P=0$.

Reduction of the $D$-dimensional Einstein--Hilbert term using these
expressions leads directly to the following Lagrangian in $d=3$:
$$
\hat{E}\hat{R}=e\bigl[R-\tr(P_aP^a)-\fr4e^{2\phi}(F_{ab}F^{ab})-D_a\phi
D^a\phi\bigr]\komma\Eqn\RAction
$$
where one should keep in mind that there is a hidden contribution to
the kinetic term of the dilaton $\phi$ in the $GL(n)$ coset term. 
Note, however,
that even after putting the two singlet terms together the kinetic
term is not conventionally normalized in our conventions; see below
for further details. The equations of motion that we will need in
the following are (in $d=3$)
$$
\eqalign{ R_{ab}&=\tr(P_aP_b)+D_a\phi D_b\phi
+\fr2e^{2\phi}\left[(F_a{}^cF_{bc})-\fr2\eta_{ab}(F^{cd}F_{cd})\right]\komma\cr
(D^aF_{ab})^i&=-(P^aF_{ab})^i-2D^a\phi F_{ab}^i \komma\cr
(D^aP_a)^{ij}&=\fr4e^{2\phi}F_{ab}^iF^{jab}\punkt\cr }\eqn
$$
Note that the equation of motion for $\phi$, 
$D^aD_a\phi=\fr4e^{2\phi}(F^{ab}F_{ab})$,  follows directly from
the last equation above since $\tr P_a=D_a\phi$. In the next
section we will apply this Ansatz to derive the compactification of
the $\hat R^2$ term.

Before leaving this review of the dimensional reduction we would
like make more explicit the relation of our conventions to the ones
in \eg\ ref. [\LambertWestI]. In that paper the Ansatz is written as
$$
\hat{E}^{a}=e^{\alpha\phi}e^{a}\komma\qquad\hat{E}^{i}
=e^{\beta\phi}(dy^{\mu}+A^{\mu})\tilde e_{\mu}{}^{i} \komma\eqn
$$
where the internal vielbein $\tilde e_\mu{}^i$ is an element of
$SL(n)$. Furthermore, the parameters $\alpha$ and $\beta$ are
determined to satisfy ${\alpha}^2=\Fr{n}{2(d-2)(n+d-2)}$ and
$\beta=-\Fr{d-2}{n}\alpha=-\sqrt{\Fr{d-2}{2n(D-2)}}$ in order for
the reduction to produce a canonical Einstein--Hilbert term and a
properly normalised kinetic term for the scalar $\phi$. In fact,
using the above Ansatz the coefficient in front of the scalar
kinetic terms reads
$$
(d-1)(d-2){\alpha}^2+2n(d-2)\alpha\beta+n(n+1)\beta^2\punkt\eqn
$$
Since our Ansatz corresponds to $d=3$, $\alpha=-1$, and
$\beta=\fr{n}$ we find the coefficient to be $1+\fr{n}$. This is
consistent with our action in eq. (\RAction) above if one extracts
the contribution to the scalar kinetic term from the coset term. The
choice $\beta=\fr{n}$ is natural, since it keeps intact the $GL(n)$
element that will be a building block of $SL(n+1)$ in the following
section. Finally, note that the field strength $F^i$ appearing in
eq. (\RAction) has an extra $\phi$ dependence hidden in the internal
vielbein.

\section\Rtwo{Toroidal dimensional reduction of $R^2$}We now
consider adding to the Einstein--Hilbert action terms of higher
order in the Riemann tensor . In the present paper, we only treat
one such deformation at the time, and think of it as the
next-to-leading term in an infinite expansion in a dimensionful
parameter formed from $\a'$ or Newton's constant.

At the level of $R^2$ there is only one possible term, modulo field
redefinitions, namely $\hat R_{ABCD}\hat R^{ABCD}$. At
next-to-leading order, field redefinitions give changes in the
action containing the lowest order field equations, so any term
containing the Ricci tensor can be thrown away without loss of
generality. The dimensional reduction (setting $A=(a,i)$ etc) will
result in an expression that contains the following kind of terms:
the square of $R_{abcd}$, two $F_{ab}^{i}$ field strengths
contracted to one $R_{abcd}$, plus $F_{ab}^{i}$, $P_{a}^{ij}$, and
$D_{a}\phi$ combined into terms with four such fields, or to
terms with three or two fields together with one or two covariant
derivatives $D_{a}$, respectively.

We note at this point that modulo field equations $\hat R_{ABCD}\hat
R^{ABCD}$ is equivalent to the Gauss--Bonnet term ${\scr
L}_{GB}=\hat E (\hat R_{ABCD}\hat R^{ABCD}-4\hat R_{AB}\hat
R^{AB}+\hat R^2)$. The fact that the integral of this expression,
$\int d^Dx{\scr L}_{GB}\sim \int\e_{A_1\ldots A_D}\hat
R^{A_1A_2}\wedge \hat R^{A_3A_4} \wedge\hat
E^{A_5}\wedge\ldots\wedge\hat E^{A_D}$, is a topological invariant
in some dimension ($D=4$) implies that it has no two-point function
(the terms quadratic in fields are total derivatives). Perhaps less
well-known is that this feature repeats itself at the level of three
fields in the scalar sector. This is an effect of the dimensional
reduction. It is quite trivial to convince oneself that any
three-point coupling $P^2DP$, modulo the lowest order field equation
(representing the freedom of field redefinitions) is a total
derivative. However, as we will discuss more later, for $R^3$ and
$R^4$ terms related to topological invariants in six and eight
dimensions, this property holds only for terms containing three and
four fields, respectively. 

To present the result of the dimensional reduction of $\hat
R_{ABCD}\hat R^{ABCD}$ it is convenient to first note that the
splitting of the indices $A=(a,i)$ etc gives
$$
\eqalign{
\hat{R}_{ABCD}\hat{R}^{ABCD}&=\hat{R}_{abcd}\hat{R}^{abcd}+4\hat{R}_{ibcd}\hat{R}^{ibcd}+
2\hat{R}_{ijcd}\hat{R}^{ijcd}\cr
&+4\hat{R}_{ibkd}\hat{R}^{ibkd}+4\hat{R}_{ijkd}\hat{R}^{ijkd}
+\hat{R}_{ijkl}\hat{R}^{ijkl}\punkt\cr} \eqn
$$
At this point we suppress the dilaton dependence in the higher
curvature terms. It should of course be kept for a complete treatment,
but will be irrelevant for the considerations in this and the
following sections. Formally, this amounts to setting $\phi=0$,
which implies $\tr P=0$.
We then get
$$
\eqalign{
\hat{R}_{abcd}\hat{R}^{abcd}&=R_{abcd}R^{abcd}-\Fr3{2}R_{abcd}(F^{ab}F^{cd})\cr
&\qquad+
\Fr3{8}\left[(F^{ab}F^{cd})(F_{ab}F_{cd})+(F^{ab}F^{cd})(F_{ac}F_{bd})\right]
\komma\cr
\hat{R}_{ibcd}\hat{R}^{ibcd}&=(D_{[c}F_{d]b}D^{c}F^{db})
-2(F_{cd}P_{b}D^{c}F^{db})+(F_{ab}P_cP^cF^{ab})\komma\cr
\hat{R}_{ijcd}\hat{R}^{ijcd}&=\fr8\left[(F_a{}^cF_{bc})(F^a{}_dF^{bd})
-(F^{ab}F^{cd})(F_{ac}F_{bd})\right]+2(F^{ae}P_{[a}P_{b]}F^b{}_{e})\cr
&\qquad-2\tr(P_{a}P_{b}P_{a}P_{b})+2\tr(P^aP_aP^bP_b)
\komma\cr
\hat{R}_{ibkd}\hat{R}^{ibkd}&=\Fr1{16}(F_a{}^cF_{bc})(F^a{}_dF^{bd})
+\tr(P^aP_aP^bP_b)-\fr2(F^a{}_{e}P_{b}P_{a}F^{be})\cr
&\qquad+
2\tr(D_{a}P_{b}P^{a}P^{b})
-\fr2(F^a{}_{e}D_{a}P_{b}F^{be})+\tr(D_{a}P_{b}D^{a}P^{b})
\komma\cr
\hat{R}_{ijkd}\hat{R}^{ijkd}&=\fr2(F^a{}_{c}F^{bc})\tr(P_{a}P_{b})
-\fr2(F^a{}_{e}P_{b}P_{a}F^{be})\komma\cr
\hat{R}_{ijkl}\hat{R}^{ijkl}
&=2\tr(P_{a}P_{b})\tr(P^aP^{b})-2\tr(P_{a}P_{b}P^{a}P^{b})
\punkt\cr } \eqn
$$
All traces and scalar products are over internal indices, all
space-time
indices are explicit.
Two of the above Riemann tensor components
depend explicitly, as well as implicitly after integration by parts,
on the field equations. After using the lowest order field equations
obtained from the reduction of the Einstein--Hilbert term, we find
that the expressions for these components become (modulo total
derivative terms and including the combinatorial factors above)
$$
\eqalign{
4\hat{R}_{ibcd}\hat{R}^{ibcd}
&=R^{abcd}(F_{ab}F_{cd})-2R^{ab}(F_{a}{}^cF_{bc})-
\fr2(F^{ab}F^{cd})(F_{ab}F_{cd})\cr
&\qquad+6(F_{a}{}^cP^{b}P^{a}F_{bc})+
2(F^{ab}P^cP_cF_{ab})\komma\cr
4\hat{R}_{ibkd}\hat{R}^{ibkd}&=-4R_{ab}\tr(P^aP^b)
+\fr4(F_a{}^cF_{bc})(F^a{}_dF^{bd})
+\fr8(F^{ab}F^{cd})(F_{ab}F_{cd})\cr
&\qquad-4\tr(P^aP_aP^bP_b)+8\tr(P_{a}P_{b}P^{a}P^{b})\cr
&\qquad-2(F^a{}_{e}P_{a}P_{b}F^{be})-2(F^{ab}P^cP_cF_{ab})\punkt\cr}
\eqn
$$
Note that we have not yet implemented the Einstein equation since it
will only produce terms with short traces, that is over two $P's$,
and these will not enter the discussion below. It is for the same
reason that we can neglect the dependence on the scalar $\phi$ in
the above formulae.
Here we have also made use of the Maurer--Cartan equations and
Bianchi identities which in the particular case of $R^2$ terms
implies that no derivatives appear anywhere (it is straightforward
to show that this is true also for non-constant $\phi$). 
As we will see in later
sections this nice feature will not occur for $R^n$ with $n>2$.

In $d=3$ the two-forms $F$ can be dualised to one-forms $f$, turning
the graviphoton degrees of freedom into scalars. Dualisation is
performed by adding a term $\int u_\m dF^\m$ to the action, thus
enforcing the Bianchi identity of $F$ with a Lagrange multiplier,
and treating $F$ as an independent field. Solving the algebraic
field equations for $F$ in terms of $du$ and reinserting the
solution into the action gives the action in terms of the scalar
dual graviphotons $u_\m$. At the level of the Einstein--Hilbert
action, reintroducing the scalar, this procedure gives the
Lagrangian
$$
{\scr L}_{\hbox{\xxrm
dual}}=e\bigl[R-\tr(P_aP^a)-\fr2(f_af^a) -D_a\phi
D^a\phi\bigr]\komma\eqn
$$
where the dualised field strength is given by $F^i=e^{-\phi}{\star}f^i$.
It has the Bianchi identity $Df+f\wedge P+f\wedge d\phi=0$ and equation
of motion $D^af_a-P^af_a-D^a\phi f_a=0$, and is obtained from the scalar as
$f=e^{-\phi}e^{-1}du$.
The dualised scalars fit together with the $GL(n)$ ones parametrising
the internal torus into an element of $SL(n+1)$ as
$$
G=\left[\matrix{e^{-\phi}&0\cr
                e^{-\phi}u&e\cr}\right]\komma\eqn
$$
which gives the $SL(n+1)$ Maurer--Cartan form
$$
{\scr P}+{\scr Q}=G^{-1}dG=\left[\matrix{
               -d\phi&0\cr
               f=e^{-\phi}e^{-1}du&e^{-1}de\cr
}\right]\punkt\eqn
$$
The $SL(n+1)$ symmetry of the dimensionally reduced Einstein--Hilbert
action is manifested as
$${\scr L}_{\hbox{\xxrm dual}}=e\bigl[R-\tr({\scr P}_a{\scr P}^a)
\bigr]\punkt\eqn
$$

We note that, at lowest order, the Lagrange multiplier term
contributes to the action (in fact, so that the kinetic term keeps
its correct sign after dualisation). When the action contains higher
order interaction terms, the equations of motion for $F$ become
non-linear, and one will get a non-linear duality relation between
$F$ and $f$. In general one has to be careful about this, but it is
straightforward to check that for any next-to-leading term, the
non-linearities cancel between the $F^2$ term and the Lagrange
multiplier term. To next-to-leading order, which is all we treat in
this paper, the correct dualised version of the higher-curvature
term is obtained by direct insertion of the linearly dualised
graviphotons.

In view of this it is of course interesting to check if the pure $P$
terms, respecting the manifest ${\frak so}(n)$ symmetry, can combine
with the graviphotonic scalars to form the enlarged symmetry ${\frak
sl}(n+1)$ also when the $R^2$ terms are included.  To this end we
collect the terms of the form $\tr(P_{a}P_{b}P^{a}P^{b})$ and
$\tr(P_{a}P^{a}P_{b}P^{b})$ together with the terms containing $F$'s
that would mix with them under ${\frak sl}(n+1)$.

The result is
$$
2\tr(P_{a}P_{b}P^{a}P^{b})+2(F_{ac}P^{b}P^{a}F_b{}^c)
\eqn
$$
(\ie, the terms $\tr(P_{a}P^{a}P_{b}P^{b})$ cancel out), which
becomes, after dualisation of the two-forms $F^{i}$ to one-forms
$f^{i}$ as discussed above,
$$
2\tr(P_{a}P_{b}P^{a}P^{b})+2(f^{a}P_{a}P^{b}f_{b})-2(f^{a}P_{b}P^{b}f_{a})
\punkt\Eqn\PPPP
$$
This should then be compared to the $SL(n+1)$-covariant
expression $\tr{\scr P}^4$. The terms contributing uniquely to this ``long
trace'', and not to $(\tr{\scr P}^2)^2$, are of the types $\tr P^4$ and
$(fPPf)$ as above, together with $\partial\phi(fPf)$, with tangent indices
placed in all possible ways.
With the parametrisation of the $SL(n+1)/SO(n+1)$ coset as above, we get
$$
\eqalign{
\tr(\P^a\P_a\P^b\P_b)&=\tr(P^aP_aP^bP_b)
    +\fr2\left[(f^aP_bP^af_b)+(f^aP^bP_bf_a)\right]\cr
    &\qquad+\fr{16}\left[(f^af_a)(f^bf_b)+(f^af^b)(f_af_b)\right]+
    \pi^a(f_aP^bf_b)\cr
    &\qquad+\fr2\left[\pi^a\pi_a(f^bf_b)+\pi^a\pi^b(f_af_b)\right]
    +\pi^a\pi_a\pi^b\pi_b\komma\cr
\tr(\P^a\P^b\P_a\P_b)&=\tr(P^aP^bP_aP_b)+(f^aP^bP_af_b)\cr
    &\qquad+\fr8(f^af^b)(f_af_b)+\pi^a(f^bP_af_b)+\pi^a\pi^b(f_af_b)
     +\pi^a\pi_a\pi^b\pi_b\komma\cr
}\Eqn\PPPPSL
$$
where $\pi=-d\phi$ is the upper left corner component of $\P$. It
seems hard to reconcile eq. (\PPPP) with a possible ${\frak
sl}(n+1)$. In fact, the coefficients of the two terms are dictated
by the $\tr P^4$ terms. Of the three structures $(fPPf)$ consistent
with $SL(n)$, only two linear combinations are allowed by $SL(n+1)$.
The terms from dimensional reduction in eq. (\PPPP) are not the ones
required by eq. (\PPPPSL).

In the above calculation, the volume factor $e^\phi$ of the internal
torus has been omitted (set to $1$). After dualisation, any term
from $\hat R^p$ carries an overall factor $e^{2(p-1)\phi}$. This
factor tells us that the terms obtained by dimensional reduction
cannot be $SL(n+1)$-invariant, since $\phi$ is one of the scalars
parametrising the coset $SL(n+1)/SO(n+1)$. Neither is this expected
from string theory or M-theory, since quantum corrections break the
global symmetry group to a discrete version. The terms obtained from
the reduction will not be the whole answer, but its large volume
limit. The torus volume factor may be obtained as the large volume
limit of an automorphic form. As we will see later, the observation
that the tensor structure does not match with $SL(n+1)$ covariance
means that scalar ($SO(n+1)$-invariant) automorphic forms (\ie,
functions) do not suffice, and calls for the introduction of
automorphic forms transforming under $SO(n+1)$. Similar conclusions
are reached in ref. [\LambertWestII] based on an investigation of the
root and weight structure of the scalar prefactors.

At this point, we could of course extend the investigation to other terms
by including $d\phi$ and considering also ``short'' traces. However, as
we already have demonstrated the need for transforming automorphic forms,
we will now show how any term obtained in the reduction can be matched
to such constructions.

\section\TransfAuto{Transforming automorphic forms}Previous work
by Green et al [\SixteenFermions] (see also [\GreenSethi]) indicates
how the apparent contradiction found in the previous section should
be resolved. In fact, as we will see in later sections, there are
also terms arising in the compactification of $R^4$ from $D=11$ to
$d=3$ that are not immediately compatible with the $SL(9)$ subgroup
of $E_{8(8)}$. We suggest that the proper interpretation of these
results is that they should be viewed as the large volume limit of
an $SL(9,\Z)$-invariant constructed from transforming automorphic
forms and non-scalar products of the fields in question. This turns
out to hold for the $R^2$ terms of the previous section on reduction
from any $D$ to $d=3$. Of course, consistency with
decompactification requires that the automorphic form, in the large
volume limit, does not diverge and has as its only remnant after
decompactification the very term that was used as starting point for
the compactification.

Appendix A describes the construction of automorphic forms, scalar
as well as transforming ones. (For a partly overlapping discussion
see the Appendix of ref. [\LambertWestII].) We now apply this
construction to the quartic terms of the previous section, although
it will be obvious that the treatment is general. For any
irreducible $SO(n+1)$ representation $r$ contained in the symmetric
product of four symmetric traceless tensors, we can form the
combination $\psi^{(r)}_{IJ,KL,MN,PQ}
\P^{aIJ}\P_a^{KL}\P^{bMN}\P_b^{PQ}$, where $\psi$ is an automorphic
form transforming in the representation $r$. The symmetric product
of four symmetric traceless $SO(n+1)$ tensors contains 23
irreducible representations for any $n\geq8$, and this is then the
number of $SL(n+1,\Z)$-invariant terms we can write down starting
from the symmetric traceless representation\foot\star{The ``weight''
of each automorphic form, as defined in appendix A, is fixed by the
overall volume factor. We ignore ambiguities from products of
automorphic forms, where only the sum of weights will be determined,
as well as from the use of different Casimirs in the sum defining
the automorphic form. Terms differing in these respects are
indistinguishable in the large volume limit.}. This is however true
only when all the indices on $\P$'s are contracted with indices on
an automorphic form constructed as in appendix A. The actual number
is larger, since nothing prevents us from taking products of such
automorphic forms and invariant tensors without symmetrising all
indices---there is no {\it a priori} reason to symmetrise in $\P$'s
with different space-time indices.

In the case of an $SL$ group, it is preferable to build automorphic
forms from the fundamental representation (although this option does
not exist if we want \eg\ $SL(9)$ as a subgroup of $E_{8(8)}$). By
using the automorphic forms built from the fundamental
representation, we have seen in Appendix A that the only surviving
part in the large volume limit is the one with all indices equal to
0 (the first component in our $SL(n+1)$ matrices) [\LambertWestII].
The part of $\hat{R}^2$ containing $P^4$ comes from an $SO(n+1)$
scalar automorphic form. Since $\P_{0i}=\fr2f_i$ and $\P_{00}=\pi$,
we can always choose to insert an even number of zeros in the
positions we like, and thereby arrange for products of transforming
automorphic forms and $SO(n+1)$-invariant tensors to have a large
volume limit reproducing any of the $SO(n)$-invariant terms
occurring in the reduction. The matching can be made recursively, in
increasing number of $0$ indices.
We take the long trace as example.
The terms with scalar automorphic forms are determined from the $\tr P^4$
terms to be proportional to $\psi^{(1)}\tr(\P^a\P^b\P_a\P_b)$, with the
notation of Appendix A.
Subtracting its large volume limit from the actual result of the
reduction, given in part by eq. (\PPPP), there is a remainder
proportional to $(f^aP_aP^bf_b)-(f^aP^bP_bf_a)-(f^aP^bP_af_b)$. This implies
a term proportional to
$\psi^{(2,1)}_{IJ}(\P^a\P_a\P^b\P_b-\P^a\P^b\P_b\P_a-\P^a\P^b\P_a\P_b)^{IJ}$.
With a more complete expansion of the reduced curvature term, it can always
be matched to the large volume limit of an
expression in terms of $\P$'s and automorphic forms.

It should of course be checked that automorphic forms exist that give
the correct power of the torus volume factor obtained from the reduction.
A term $\hat R^p$ gives an overall factor $e^{2(p-1)\phi}$. Suppose we
try to obtain some corresponding terms with a product of $M$ automorphic
forms, each with $2l_k$ fundamental indices and weight $w_k$,
$k=1,\ldots,M$.
Convergence of the sum defining the automorphic forms demands $2(w_k-l_k)>1$,
and the large volume limit will yield a dilaton dependence
$\exp({\sum_{k=1}^M2(w_k-l_k)\phi})$. Matching with the reduction gives
$p-1=\sum_{k=1}^M(w_k-l_k)$ and thus $M<2(p-1)$. The $\hat R^2$ term gives
room for one automorphic form, which is exactly what we need.

It is not clear what to expect for this kind of symmetry enhancement
to $SL(n+1)$ in the context of $\hat R^2$ terms. There are \eg\ no known
examples from string/M-theory that make use of such a step, so the
$\hat R^2$ term should probably be seen as a toy model to set the
framework for the higher-curvature terms. (The heterotic string has
an $\hat R^2$ term, but the symmetry enhancement is to $SO(n,n)$; this
case is treated in ref. [\Meissner].) The situation is different for
the $\hat R^4$ terms, which are the first higher derivative terms to
arise in the maximally supersymmetric string theories in 10
dimensions, as well as in 11-dimensional M-theory. When compactified
to three dimensions all degrees of freedom are collected into a
coset based on $E_{8(8)}$ when starting from a two-derivative
action. We now proceed to discuss $\hat R^3$ and $\hat R^4$ terms in this
context with the goal of understanding the role of $SL(n+1)$, and to
develop methods that might eventually be useful in dealing with the
more complicated case of $E_{8(8)}$.

Automorphic forms of $SL(2,\Z)$ transforming under $U(1)$ have been encountered
in loop calculations with external fermions in string theory compactified
on a circle to $d=9$ [\SixteenFermions]. We expect that the appearance of
transforming automorphic forms is generic in a situation where the external
fields of the diagram transforms under $K$. In the specific case in
ref. [\SixteenFermions], the contribution was shown to disappear in the
M-theoretic large volume limit. We note that this limit is a quite different
one in terms of the $SL$ group involved than in our case. We deal with
a larger symmetry $SL(n+1)$ appearing after dualisation of the graviphotons,
and blow up a certain parameter, that, had our $SL(n+1)$ element been a
vielbein on $T^{n+1}$, would have corresponded to shrinking one direction and
consequently blowing up the other $n$ directions. In the $SL(2)$ case, the
large volume limit has nothing to do with the $SL(2)$ element parametrising
the shape of $T^2$, but with the determinant of a $GL(2)$ element blowing up.
We have shown how to match combinations of $\P$'s and automorphic forms that
are designed to survive in the large volume limit. It would be interesting
to compare such a construction (not for terms corresponding to $\hat R^2$, but
presumably to $\hat R^4$) to actual loop calculations.

\section\Rthree{The case of $R^3$}Our main concern in the rest of the paper
is the investigation of the $\hat R^4$ terms which are part of the first
non-trivial correction in M-theory and type \II\ string theory.
Before doing that we would however like to emphasize some aspects of
$\hat R^3$ terms. The $\hat R^2$ 
terms of section \Rtwo\ were a testing ground for
the ideas but turned out to have some special non-generic features,
such as the effective vanishing of all terms with second derivatives
on scalar fields. As we will see below this feature it not found for
$\hat R^3$ and higher terms. Here we also take the opportunity to
introduce some diagrammatic methods that will be tremendously
helpful in keeping track of index structures of increasing
complexity as we go to higher powers of the Riemann
tensor.

Again, the $\hat R^3$ terms are seen as a next-to-leading order
correction to the Einstein--Hilbert action (\ie, there are no $\hat R^2$
terms). Any term which contains lowest order field equation can be
removed by a field redefinition, so we leave them out from the
start. We thus want to list all possible terms where indices are
contracted between different Riemann tensors. We represent each
contracted index by a line, and each Riemann tensor by the endpoints
of four such lines. The lines whose endpoints meet represent an
antisymmetric pair of indices. The sign is fixed by letting the
indices, as they sit on $\hat R$, run in the clockwise direction in
the diagram. The only structure not accounted for is $\hat
R_{A[BCD]}=0$, which has a simple diagrammatic expression.

A basis for the two inequivalent $R^3$ terms can be taken as
\vskip3\parskip
\noindent
\epsfxsize=1.5cm
\raise.65cm\hbox{(1):} \epsffile{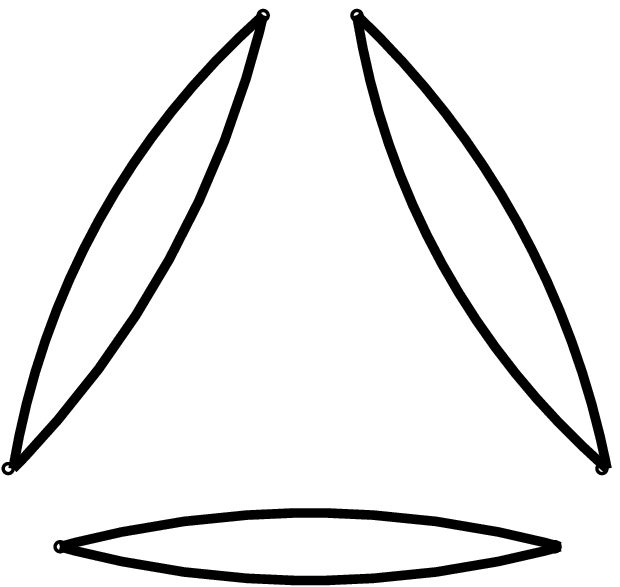}
\hskip5mm
\epsfxsize=1.5cm
\raise.65cm\hbox{(2):} \epsffile{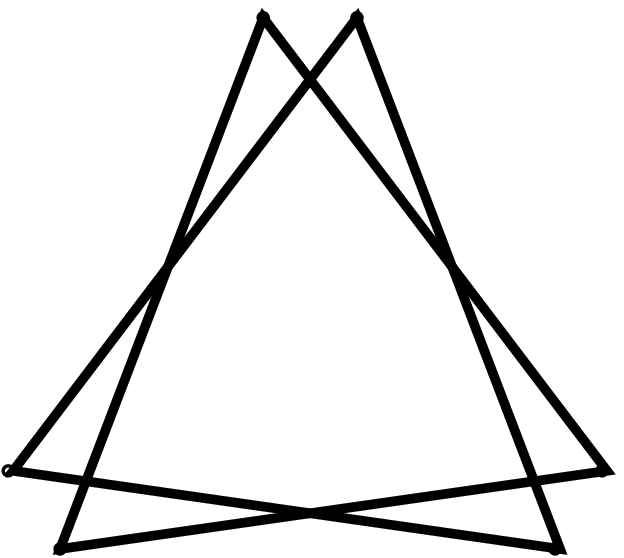}

\noindent thus representing $\hat R_{AB}{}^{CD}\hat
R_{DC}{}^{EF}\hat R_{FE}{}^{BA}$ and $\hat R_A{}^B{}_D{}^E\hat
R_B{}^C{}_E{}^F\hat R_C{}^A{}_F{}^D$, respectively. One may also
consider the contraction $\rrrdiag3$, but it is related to the
ones in the basis, using the Bianchi identity $\hat R_{A[BCD]}=0$,
as $\rrrdiag3=\fr4(1)+(2)$.

At this level, there is one obvious combination that does not give any
three-point couplings. This is the ``Gauss--Bonnet'' term,
$$
\eqalign{ \e\e\hat R^3&=\e^{A_1\ldots A_D}\e_{B_1\ldots B_D} \hat
R_{A_1A_2}{}^{B_1B_2}\hat R_{A_3A_4}{}^{B_3B_4} \hat
R_{A_5A_6}{}^{B_5B_6} \d_{A_7\ldots A_D}^{B_7\ldots B_D}\cr
&=32(D-6)!\{(1)+2(2)\}\cr} \komma\Eqn\GaussBonnetThree
$$
which is topological in $D=6$ and lacks three-point couplings in any
dimension.
The general form of the scalar terms will be
$(D\P)^3+\P^2(D\P)^2+\P^4D\P+\P^6$, where ``$\P$'' denotes any of
$P$, $f$ and $\partial\phi$, but we are guaranteed that the first
term vanish for this specific combination.

To see this explicitly, and to derive further properties
relying on the dimensional reduction, we concentrate on the pure 
$P$ terms (note that this truncation is consistent and implies
$\tr P=0$). They are extracted in the
Riemann tensor derived in section 2:
$$
\eqalign{ \hat{R}_{ab}{}^{cd}&=-4\delta_{[a}^{[c}\tr(P_{b]}P^{d]}) +
\delta_{[a}^c\delta_{b]}^d\tr(P_eP^e)\komma\cr
\hat{R}_{ab}{}^{cl}&=0\komma\cr
\hat{R}_{ab}{}^{kl}&=-2(P_{[a}P_{b]})^{kl}\komma\cr
\hat{R}_{aj}{}^{cl}&=
           -(D_aP^c)_j{}^l
           -(P_aP^c)_j{}^l\komma\cr
\hat{R}_{aj}{}^{kl}&=0\komma\cr
\hat{R}_{ij}{}^{kl}&=-2(P_e){}_{[i}{}^k(P^e){}_{j]}{}^l
 \punkt\cr
 }\Eqn\RiemannPureP
$$
Notice that Einstein's equations in three dimensions have been
used to obtain the specific form of $\hat{R}_{ab}{}^{cd}$.

The two independent cubic contractions of the Riemann tensor
components above become after compactification, and keeping only
terms which give pure $P$ contributions\foot\star{The combinatorial 
factors are easily read off from the diagrams. Splitting 
of the indices in two classes, space-time and internal, corresponds to 
coloring the lines in the diagrams with two colors. The factors are given
by the number of ways this can be done.},
$$
\eqalign{&\hat R_{AB}{}^{CD}\hat R_{DC}{}^{EF}\hat
R_{FE}{}^{BA}\cr
&\qquad=R_{ij}{}^{kl}R_{lk}{}^{mn}R_{nm}{}^{ji}+3R_{ab}{}^{kl}
R_{lk}{}^{mn}R_{nm}{}^{ba}
+8R_{aj}{}^{cl}R_{lc}{}^{en}R_{ne}{}^{ja}\cr
&\qquad+R_{ab}{}^{cd}R_{dc}{}^{ef}R_{fe}{}^{ba}
+3R_{ab}{}^{cd}R_{dc}{}^{ij}R_{ji}{}^{ba}\cr}\Eqn\RCubeOne
$$
and
$$
\eqalign{ &\hat R_A{}^B{}_D{}^E \hat R_B{}^C{}_E{}^F\hat
R_C{}^A{}_F{}^D\cr 
&\qquad=R_i{}^j{}_l{}^m R_j{}^k{}_m{}^n
R_k{}^i{}_n{}^l+3R_a{}^j{}_d{}^m R_j{}^k{}_m{}^n R_k{}^a{}_n{}^d+
2R_i{}^j{}_d{}^e R_j{}^k{}_e{}^f R_k{}^i{}_f{}^d\cr
&\qquad+6R_i{}^b{}_d{}^m R_b{}^k{}_m{}^f R_k{}^i{}_f{}^d+R_a{}^b{}_e{}^f
R_b{}^c{}_f{}^gR_c{}^d{}_g{}^e+3R_a{}^b{}_c{}^d
R_b{}^i{}_d{}^jR_i{}^a{}_j{}^c\punkt\cr}\Eqn\RCubeTwo
$$
We now insert the $P$-dependent terms from above. For the
purposes here it is sufficient to collect only the $(DP)^3$ and
$P^2(DP)^2$ terms, while remembering that $\tr P=0$.

For $(DP)^3$, which gets contribution entirely from $(1)$\foot\dagger{This
term comes only from $\ss R_{\sss aj}{}^{\sss cl}$, which means that it
gets contributions only from colored diagrams with alternating color on
all cycles. A diagram containing a cycle with an odd number of lines 
can not contribute.}, 
it is
straightforward to show that it is a total derivative
$$
(DP)^3=\tr(S_a{}^b{}S_b{}^c{}S_c{}^a{})=D_c\left[
\tr(P^aS_a{}^b{}S_b{}^c{}) - \fr2 \tr(P^cS_a{}^b{}S_b{}^a{}) \right]
\Eqn\DPCube
$$
(as always, modulo lowest order equations of motion),
where $S_{ab}=D_aP_b$. The tensor $(S_{ab})_{ij}$ is symmetric in
both $(ab)$ and $(ij)$, and $S_a{}^a{}$ is the kinetic term in the
equation of motion for $P_a$. The fact that, modulo equations of
motion, the $(DP)^3$ term is a total derivative is expected for the
highest derivative term in a Gauss--Bonnet combination of any order,
but we see that after dimensional reduction the scalar three-point couplings
vanish for any $\hat R^3$ term.

Doing a similar analysis for the $(DP)^2 P^2$ terms, there are 10
algebraically independent structures
$$
\eqalign{ 
(\i)&=\tr(S_a{}^bS_b{}^aP_cP^c)\komma\cr
(\ii)&=\tr(S_a{}^bS_b{}^cP_cP^a)\komma\cr
(\iii)&=\tr(S_a{}^bS_b{}^cP^aP_c)\komma\cr
(\iv)&=\tr(S_a{}^bP_cS_b{}^aP^c)\komma\cr
(\v)&=\tr(S_a{}^bP_bS^a{}_cP^c)\komma\cr
(\vi)&=\tr(S_a{}^bS_b{}^a)tr(P_{c}P^{c})\komma\cr
(\vii)&=\tr(S_a{}^bS_b{}^c)tr(P_{c}P^{a})\komma\cr
(\viii)&=\tr(S_a{}^bP_{c})tr(S_b{}^aP^c)\komma\cr
(\ix)&=\tr(S_a{}^bP_{c})tr(S_c{}^aP_{b})\komma\cr
(\x)&=\tr(S_a{}^bP_{b})tr(S_c{}^aP^{c})\punkt\cr
}\Eqn\SPTwoStructures
$$
Since $(\i)$--$(\v)$ will not mix with $(\vi)$--$(\x)$, we will
consider the two groups separately. For the single trace terms, neglecting
the equations of motion, the combination
$x_1(\i)+2x_2(\ii)-(x_2-2x_3)(\iii)+x_3(\iv)+(2x_1-x_2)(\v)$ is a
total derivative for arbitrary values of $\{x_*\}$. Correspondingly
a total derivative consisting of the double trace terms must be
written as
$y_1(\vi)+(y_2+y_3)(\vii)+y_2(\viii)+(y_2-2y_3)(\ix)+(2y_1+y_3)(\x)$
for arbitrary values of $\{y_*\}$. Extracting the pure $(DP)^2P^2$
terms from (\RCubeOne) and (\RCubeTwo) we find that
$$
(1)+z(2) = 6(4-z)(\ii)+6z(\iii)+3z(\iv)  + z\left[
-\Fr3{2}(\vi)+6(\vii)-3(\viii) \right]\komma\Eqn\RCubeComb
$$
with an arbitrary parameter $z$. For the single trace terms in
eq. (\RCubeComb) $z=4$ is the only choice where they can form a total
derivative, this corresponds to the case $x_1=x_2=0$, $x_3=12$,
(which is not the Gauss--Bonnet combination from
eq. (\GaussBonnetThree)). 
For the double trace terms
in eq. (\RCubeComb), however, no choice of $z$ can make them a total derivative.
We have thus shown that the $(DP)^2P^2$ cannot vanish by partial
integrations. Unlike in the $\hat R^2$ terms, derivatives of
Maurer--Cartan forms necessarily appear. 

A more complete treatment should include also the other fields in
$\P$. One should also continue with terms of the types $\P^4D\P$ and
$\P^6$. This would imply quite some work which we do not find
motivated for $\hat R^3$. In order to access the complete expressions,
care has to be taken when using partial integrations, since terms which a
certain number of derivatives contributes to terms with fewer
derivatives via equations of motion, Bianchi identities 
and curvatures ($R$ and $F_Q$).

\section\Rfour{$R^4$ terms}In this section we start the analysis of
the $\hat R^4$ terms by presenting the 
content of  $\ttR$ and
$\eeR$ in terms of an explicitly given basis of seven elements. That
this basis is seven-dimensional is well-known [\Fulling]. We then
concentrate on the terms that after the dimensional reduction
contain only the coset variable $P_a^{ij}$. These are of the types $(DP)^4$,
$P^2(DP)^3$, $P^4(DP)^2$, $P^6(DP)$, and $P^8$. A test of the possible
r\^ole of the octic invariant of $E_{8(8)}$ derived in ref. 
[\CederwallPalmkvist] is spelt out (for details see Appendix B).
This would involve the $P^8$ terms and be rather lengthy. For that
reason we turn in section to the much simpler terms $(DP)^4$ from
which we able to draw the conclusions we are looking for.

Using the same diagrammatic notation as in the previous section,
a basis for the seven $\hat R^4$ 
terms can be taken as

\vskip3\parskip
\noindent
\epsfxsize=1.5cm
\raise.65cm\hbox{(1):} \epsffile{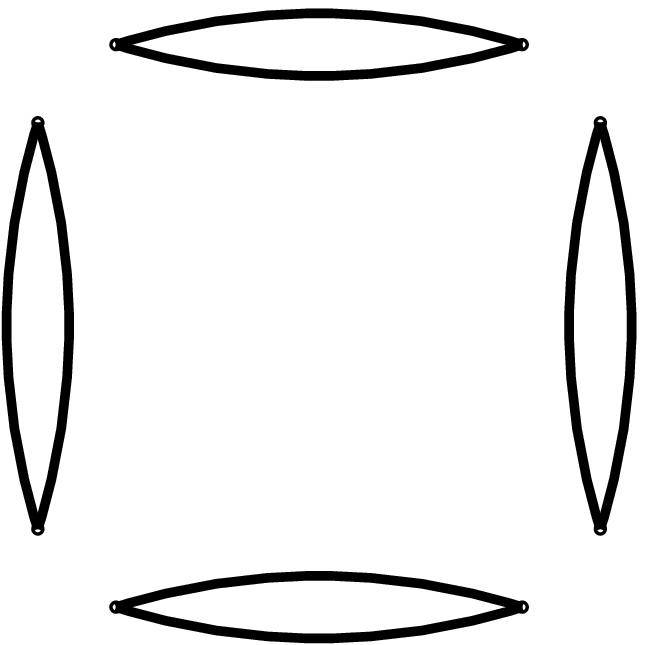}
\hskip5mm
\epsfxsize=1.5cm
\raise.65cm\hbox{(2):} \epsffile{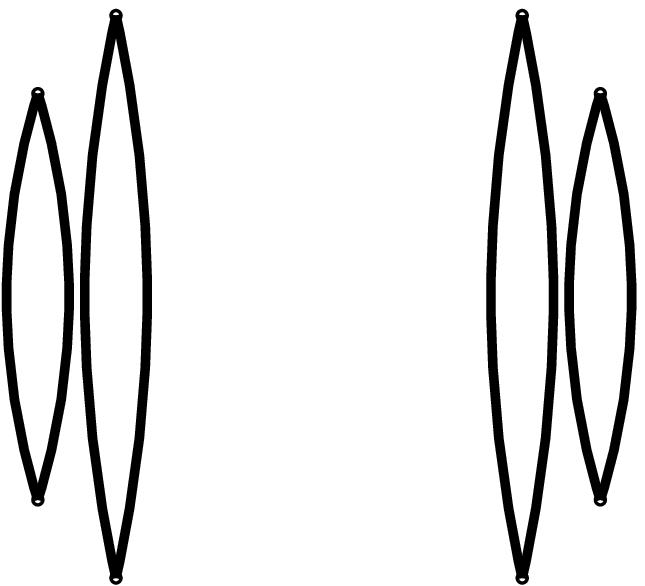}
\hskip5mm
\epsfxsize=1.5cm
\raise.65cm\hbox{(3):} \epsffile{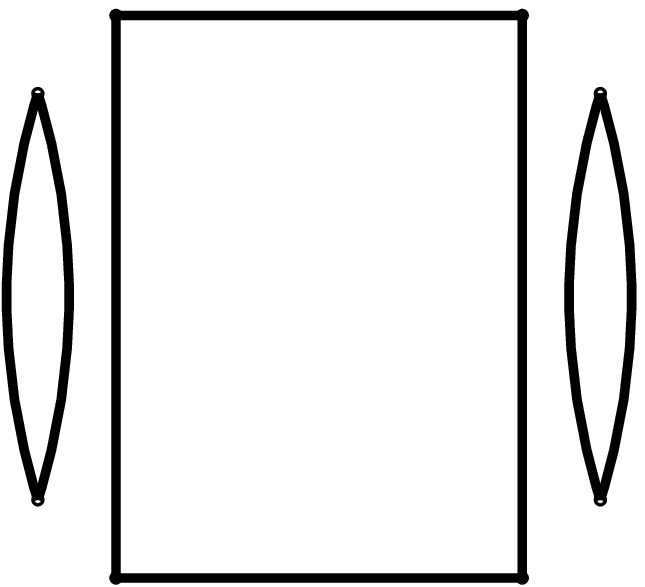}
\hskip5mm
\epsfxsize=1.5cm
\raise.65cm\hbox{(4):} \epsffile{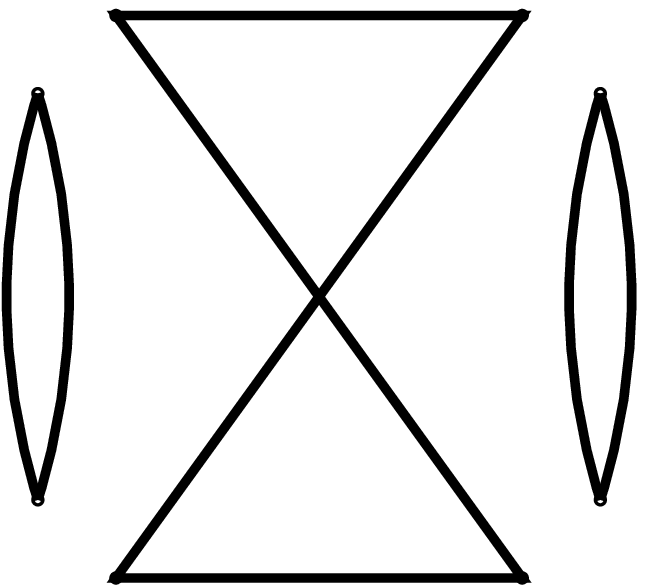}

\vskip3\parskip\noindent
\epsfxsize=1.5cm
\raise.65cm\hbox{(5):} \epsffile{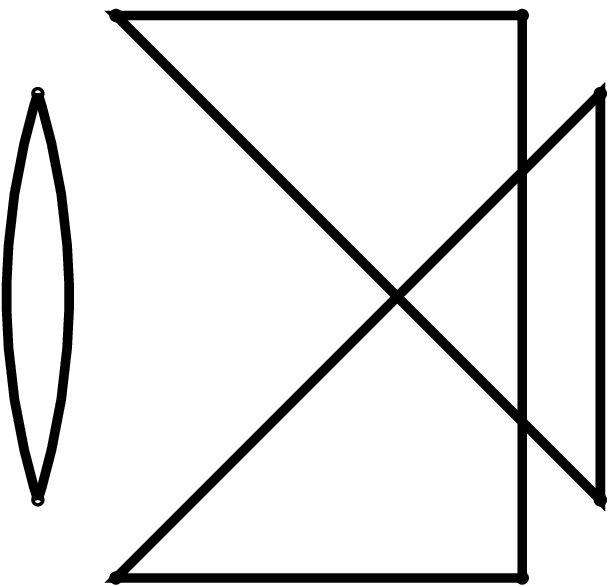}
\hskip5mm
\epsfxsize=1.5cm
\raise.65cm\hbox{(6):} \epsffile{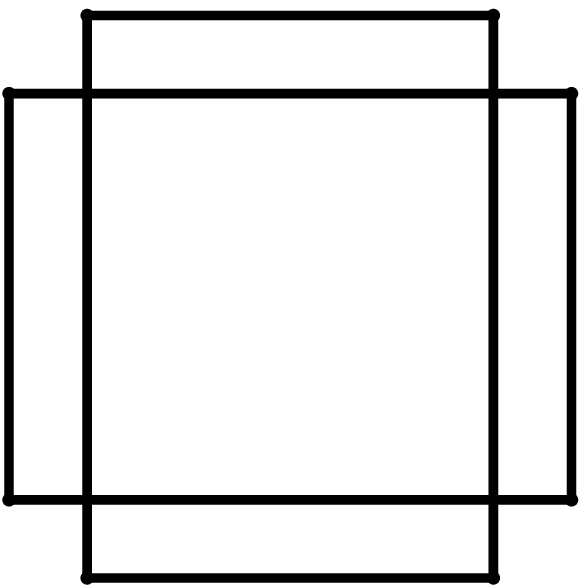}
\hskip5mm
\epsfxsize=1.5cm
\raise.65cm\hbox{(7):} \epsffile{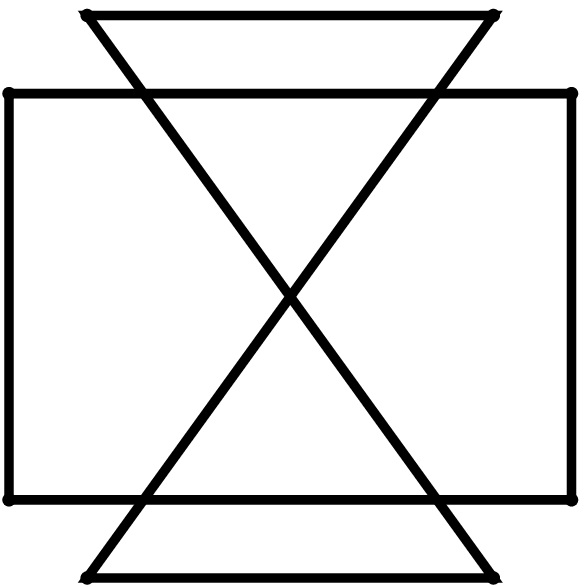}

A contraction that also occurs naturally (\eg\ in $\e\e R^4$) is
$\rdiag8$,
and it can be related to the others by using $\hat R_{[ABC]D}=0$ as follows:
Cycling on $\rdiag8$ gives
$\rdiag8=\rdiag{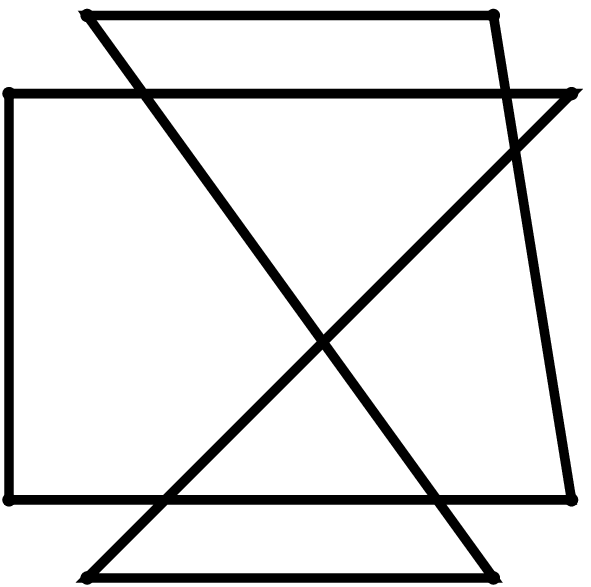}-\fr2\rdiag{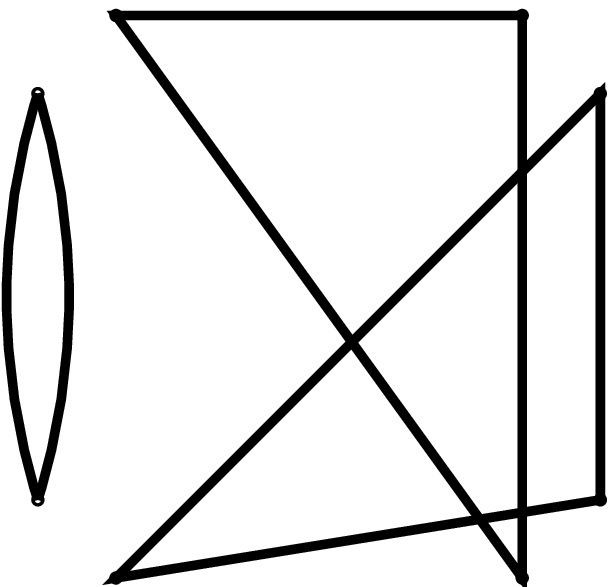}$. Cycling on $(5)=\rdiag5$ gives
$\rdiag5=\rdiag{332}+\fr2\rdiag4$, and on $(7)=\rdiag7$ gives
$\rdiag7=\rdiag{88}+\fr2\rdiag5$. Eliminating the diagrams not present in the
basis, $\rdiag{88}$ and $\rdiag{332}$, gives the relation
$\rdiag8=\fr4(4)-(5)+(7)$.

In $D=10$ and 11, the structures
$$
\eeR=\e^{A_1\ldots A_D}\e_{B_1\ldots B_D} \hat
R_{A_1A_2}{}^{B_1B_2}\hat R_{A_3A_4}{}^{B_3B_4} \hat
R_{A_5A_6}{}^{B_5B_6}\hat R_{A_7A_8}{}^{B_7B_8} \d_{A_9\ldots
A_D}^{B_9\ldots B_D} \eqn
$$
(the ``Gauss--Bonnet term'') and $\ttR$ are of special interest,
since they, or combinations of them, are dictated by string theory
calculations and by supersymmetry; see for instance the explicit
evaluation in Appendix B2 of ref. [\PetersVanhoveWesterberg] of the
appropriate superspace term given in ref. [\NilssonTollstenIII]. The
invariant tensor $t_8^{A_1A_2,A_3A_4,A_5A_6,A_7A_8}$ is defined to be
antisymmetric in the indices composing the pairs and symmetric in
the four pairs. When contracted with the an antisymmetric matrix
$M$, it is defined to give
$$
t_8^{A_1A_2,A_3A_4,A_5A_6,A_7A_8}M_{A_1A_2}M_{A_3A_4}M_{A_5A_6}M_{A_7A_8}
=24\tr M^4-6(\tr M^2)^2\punkt\eqn
$$ 
In $\ttR$, the indices are contracted
according to
$$
\eqalign{
\ttR&=t_8^{A_1A_2,A_3A_4,A_5A_6,A_7A_8}t_{8\,B_1B_2,B_3B_4,B_5B_6,B_7B_8}\cr
&\qquad\times
\hat R_{A_1A_2}{}^{B_1B_2}\hat R_{A_3A_4}{}^{B_3B_4}
\hat R_{A_5A_6}{}^{B_5B_6}\hat R_{A_7A_8}{}^{B_7B_8}
\punkt\cr}
\eqn
$$
Direct evaluation gives, in $D$ dimensions,
$$
\eqalign{
\Fr1{12}\ttR&=2(1)+(2)-16(3)-8(4)\,\,\,\phantom{+32(5)}+16(6)+32(7)\cr
-\Fr1{48(D-8)!}\eeR&=2(1)+(2)-16(3)\,\,\,\phantom{+8(4)}+32(5)+16(6)\,\,\,\phantom{+32(7)}
-32\,\rdiag8\cr }\Eqn\ttRandeeR
$$
or, with $\rdiag8$ expressed in the basis as above,
$$
\eqalign{
\Fr1{12}\ttR&=2(1)+(2)-16(3)-8(4)\,\,\,\phantom{+64(5)}+16(6)+32(7)\cr
-\Fr1{48(D-8)!}\eeR&=2(1)+(2)-16(3)-8(4)+64(5)+16(6)-32(7)\cr
}\Eqn\ttRandeeRtwo
$$
These expressions agree with the ones in \eg\ ref.
[\HyakutakeOgushi], where the basis
$\{A_1=(2),A_2=(3),A_3=(1),A_4=(4),A_5=-\rdiag{332}=\fr2(4)-(5),
A_6=(6),A_7=\rdiag8=\fr4(4)-(5)+(7)\}$ is used.

The $P^8$ terms obtained when compactifying from eleven dimensions
to three will of course form a scalar of $SO(8)$. Assuming that
these terms combine to a scalar also of the $Spin(16)/\Z_2$ that is
associated with coset $E_{8(8)}/(Spin(16)/\Z_2)$ arising in the
two-derivative sector of M-theory, the only invariant possible 
(apart from the fourth power of the quadratic one) would
be the octic invariant constructed in ref. [\CederwallPalmkvist]. As explained
in Appendix B, when reducing this to an invariant of $SO(9)$ one
finds a certain polynomial in the $SL(9)/SO(9)$ coset element that
if valid puts severe restrictions on the structure of the $P^8$
terms. However, checking this is lengthy and instead we turn to the
$(DP)^4$ terms where, as we will see below, some qualitative results we are
looking for can be obtained with much less effort.

Thus, we now concentrate on the four-point couplings, which
consequently have four derivatives. Assume, for the moment, that
$E_{8(8)}(\Z)$ invariance were to be achieved with a scalar
automorphic form. Since $E_8$ has no invariant of order four other
than the square of the quadratic Casimir (and thus the only ${\frak
so}(16)$ invariant quartic in spinors is the square of the quadratic
one), we would get the restriction that any trace $\tr(DP)^4$ has to
vanish, since this ${\frak so}(8)$ invariant cannot be lifted via
${\frak so}(9)$ to ${\frak so}(16)$.

Using the Riemann tensor with only $P$ terms (see the previous
section), it is not very difficult to derive the $(DP)^4$-terms from
the diagrams (1)-(7). Since we only want contributions with the
components $R_{aibj}$, one gets one contribution from each coloring
with two colors (for space-time and internal indices) of the graphs,
where the two colors alternate on every cycle. It follows directly
that any diagram with a cycle of odd length does not contribute.
There are none in the basis, but in the process of cycling above we
encountered the contraction $\rdiag{332}=(5)-\fr2(4)$ that then does
not contribute to $(DP)^4$.

There are 8 algebraically independent structures containing $(DP)^4$.
We enumerate them as
$$
\eqalign{
(\i)&=\tr(S_{ab}S^{ab}S_{cd}S^{cd})\komma\cr
(\ii)&=\tr(S_{ab}S_{cd}S^{ab}S^{cd})\komma\cr
(\iii)&=\tr(S_{ab}S^{bc}S_{cd}S^{da})\komma\cr
(\iv)&=\tr(S_{ab}S_{cd}S^{ac}S^{bd})\komma\cr
(\v)&=\tr(S_{ab}S^{ab})\tr(S_{cd}S^{cd})\komma\cr
(\vi)&=\tr(S_{ab}S^{cd})\tr(S_{ab}S^{cd})\komma\cr
(\vii)&=\tr(S_{ac}S^{bc})\tr(S^{ad}S_{bd})\komma\cr
(\viii)&=\tr(S_{ab}S_{cd})\tr(S^{ac}S^{bd})\komma\cr
}\Eqn\SFourStructures
$$
where $S_{ab}=D_aP_b$. One also has to take total derivatives into
account. This can be done by writing out all possible terms
$(PS^3)_a$ (there are 12) and take the divergence. As long as we only
consider $(DP)^4$, we let $S^a{}_a\rightarrow0$ and $S_{[ab]}\rightarrow0$. 
It turns out that
only two combinations of these do not produce terms $P(DP)^2D^2P$ (the
second derivative of $P$ can again be considered as symmetric and traceless),
and they lead to the combinations $(\i)+\fr2(\ii)-(\iii)-2(\iv)$ and
$\fr2(\v)+(\vi)-2(\vii)-(\viii)$ being total derivatives. (These in fact
arise from
$\tr(P_{[a}S_{b}{}^{b}S_{c}{}^{c}S_{d]}{}^{d})$ and
$\tr(P_{[a}S_{b}{}^{b})\tr(S_{c}{}^{c}S_{d]}{}^{d})$, where the antisymmetry, by
the Bianchi identity, prevents $P(DP)^2D^2P$ from arising. The counting also
holds for reduction to $d=3$, but with the combinations being total derivatives
in higher dimensions now being identically zero.)

Evaluating the contributions to the 4-point couplings from the terms
$(1),\ldots,(7)$ then gives
$$
\eqalign{
(1)&\longrightarrow16(\iii)\komma\cr
(2)&\longrightarrow16(\v)\komma\cr
(3)&\longrightarrow4(\i)+4(\vii)\komma\cr
(4)&\longrightarrow8(\iv)\komma\cr
(5)&\longrightarrow4(\iv)\komma\cr
(6)&\longrightarrow2(\iii)+2(\vi)\komma\cr
(7)&\longrightarrow(\ii)+2(\iv)+(\viii)\punkt\cr
}\eqn
$$
Demanding that the contribution vanishes, modulo total derivatives, tells
that the $R^4$ term is proportional to
$2(1)+(2)-16(3)+x(4)+(48-2x)(5)+16(6)-32(7)$ for some number $x$.
$\eeR$ (of course)
passes the test, but $\ttR$ does not. The combination $(4)-2(5)$
does, as seen above. In this calculation, $\ttR$ does not even
contribute with $(\tr S^2)^2$ terms only, as would be demanded from $E_8$
invariance. (The condition that the long contractions vanish can be expressed
as conditions on the coefficients in front of (5), (6) and (7), given
the ones in front of (1)-(4). The latter are identical in $\ttR$ and $\eeR$.)
 The ``difference'' between $\ttR$ and $\eeR$ (with the
normalisations above) is another very simple expression,
proportional to $(7)-(5)$ or to $\rdiag8-\fr4(4)$, whose
contribution to the long contractions is $(\ii)-2(\iv)\neq0$.
In conclusion, if the term $\ttR$ is present, there are four-point couplings
not only in the gravity sector but also in the scalar sector. The term
$\ttR$ can not be obtained without transforming $E_8$ automorphic forms.

We thus find a contradiction with $E_8$ unless transforming
automorphic forms are introduced. The fact that $E_8$ does
not have primitive fourth order invariant means that the $SL(8)$
invariant $D^4P^4$ terms derived here must come from an $E_8$ term
which is a double trace. Since we find non-zero single trace terms
it means that the enhanced symmetries do not generalise to higher
derivative terms obtained through compactification as described
here with scalar automorphic forms.

Given that the number of automorphic forms of $E_8$ is smaller than
that of $SL(9)$, for the same number of ${\frak so}(16)$ spinors
or ${\frak so}(9)$
symmetric traceless tensors (see appendix A), it seems reasonable to
believe that $E_8$ puts some constraints on the possible terms obtained
by reduction of pure gravity. Checking this would require
more concrete knowledge of
$E_8$ automorphic forms and their large volume limit, as well as
(presumably) a much more complete expansion of the seven $R^4$ terms.
It is not at all clear to what degree $E_8$ will single out some specific
combination of these.

Performing a loop calculation with external scalars analogous to the
ones in refs. [\GreenGutVan,\SixteenFermions] would give information on
what kind automorphic functions actually appear in an M-theory context
(although such a calculation leaves out
non-perturbative information from winding membranes and five-branes).

\appendix{Automorphic forms}Consider an element $g\in G$, where
$G$ is a Lie group. In the context of the supergravities (or sigma
models) we are considering, $g$ represents the scalar degrees of
freedom. These belong to a coset $G/K$, where $K$ is a subgroup of
$G$. In all cases under consideration, $G$ has the split (maximally
non-compact) real form, and $K$ is the maximal compact subgroup of
$G$. The coset is realised by gauging the local right action of $K$,
$g\rightarrow gk$, $k\in K$. This still leaves room for a global
left action of $G$ on $g$, $g\rightarrow\g g$, $\g\in G$. These
global $G$ transformations are however symmetries only of the
undeformed supergravities or sigma models, and are broken by quantum
effect in string theory. Higher-derivative corrections to effective
actions in string theory are expected to break $G$ to a discrete
duality subgroup $G(\Z)$, and the correct moduli space for the
scalars is not $G/K$ but $G(\Z)\backslash G/K$.

The definition of $G(\Z)$ has to be clear, of course. If $G$ is a
classical matrix group, it can be defined as the group of elements
in $G$ with integer entries in the fundamental representation. For
exceptional groups, care has to be taken to choose the relevant
discrete subgroup. Ref. [\MizoguchiSchroder] gives a definition of
$G(\Z)$ in terms of generators of the Lie algebra ${\frak g}$ of $G$
in the Chevalley basis (see also ref. [\ObersPiolineU]).
In the following it will be understood that
$G(\Z)$ is the discrete duality group relevant to M-theory
compactifications, although the construction in principle holds also
for other discrete subgroups of $G$.

The general method for building automorphic forms 
[\KazhdanPiolineWaldron,\ObersPioline,\GreenGutVan,\GreenSethi,\LambertWestII]
is to combine $g$
with some element in the discrete group (or a representation of it)
so that the resulting entity only transforms under $K$, in the sense
defined below. The invariance under $G(\Z)$ is then obtained by
summation over $G(\Z)$ (or some representation). Let $g\in G$ and
$\mu\in G(\Z)$, with the transformation rules under $G(\Z)\times K$
with group element $\g\otimes k$: $g\rightarrow\g gk$,
$\mu\rightarrow\g\m\g^{-1}$. If one forms $g^{-1}\mu g$, it
transforms as $g^{-1}\m g\rightarrow k^{-1}(g^{-1}\m g)k$, \ie, only
under $K$. One may then $K$-covariantly project $g^{-1}\mu g$ on the
representation ``${\frak g}/{\frak k}$'', \ie, the complement to
${\frak k}$ in the Lie algebra ${\frak g}$, which forms a
representation of $K$.\foot\star{This projection may be performed
by letting {\eightmath g} and {\eightmath\char22}
be represented as matrices in any faithful
representation of {\eightmath G}, the result of course being independent of the
choice of representation.} We denote the obtained building block
$\G=\Pi_{\gg/\kk}(g^{-1}\m g)$. When using tensor notation, we write
$\G_\a$, inspired by the ${\frak so}(16)$ spinor index carried by the
tangent space to $E_8{(8)}/(Spin(16)/\Z_2)$.

Let us start with the simplest kind of automorphic forms, the scalar
ones. In order for the function not to transform under $K$, we need
to form scalars from a number of $\G$'s. This is
straightforward---the algebraically independent polynomial
invariants have the same number and degree of homogeneity as the
Casimir operators of $\gg$. In fact, as observed in ref.
[\CederwallPalmkvist], they are simply the restrictions of the
Casimir operators to $\gg/\kk$. Let us denote them $C_i(\G)$,
$i=1,\ldots,r$, $r$ being the rank of $\gg$. Finally, in order to
achieve invariance under $G(\Z)$, one has to form some function of
the $C_i$'s and sum over the discrete group element $\m$. The
function should be conveniently formed so that the sum converges,
\eg\ a power function. For some ``weight'' $w$, we thus define
$$
\phi^{(i,w)}(g)=\sum_{\m\in G(\Z)}[C_i(\G(\m,g))]^{-w} \punkt
\Eqn\AutomorphicForm
$$
This automorphic function is clearly a function on the double coset
$G(\Z)\backslash G/K$.

The construction above is entirely based on $\G$, which is obtained
as (a projection of) the action of $g$ by conjugation on a discrete
group element $\m$.
Alternatively, one may start from some
representation. Especially, when $G$ is a classical matrix group, it
is simpler to let $m$ lie in the fundamental module (a row vector
with integer entries) [\ObersPioline].
Consider the case $G=SL(n)$ with $K=SO(n)$.
We form $mg$, which if $m$ transforms as $m\rightarrow m\g^{-1}$
transforms under $G(\Z)\times K$ as $mg\rightarrow (mg)k$. Then,
$mg$ is taken as a building block, and one forms the invariant
$\vert mg\vert^2=(mg)(mg)^t$. The automorphic function is
$$
\psi^{(w)}(g)=\sum_{\Z^n\backslash 0}\vert mg\vert^{-2w}\punkt\Eqn\SLAuto
$$
This construction has the advantage that the summation is easier to
perform than the one over the discrete group, but it is not
available for exceptional groups $G$. Choosing other modules yields
algebraically independent automorphic functions, as long as these
modules are formed by anti-symmetrisation from the fundamental one.
One gets again a number of functions equating the rank.

We expect that the summation in the defining equation (\AutomorphicForm),
which is over a single orbit of the discrete group, namely the group itself,
can be lifted to the summation over a lattice, quite analogously to
how the summation in eq. (\SLAuto) can be decomposed into an infinite
number of orbits. Such a lattice summation might make even automorphic
forms of exceptional groups reasonable to handle. Eq. (\AutomorphicForm),
with the replacement of the discrete group by a lattice, is well suited
for the bosonic degrees of freedom of the sigma model obtained by dimensional
reduction, since the object $\Gamma$ carries the same index structure as $P$.
When it comes to fermions, these transform under another representation
which is (an enlargement of) a spinor representation of ${\frak so}(n)$, and it
seems natural to consider spinorial automorphic forms.

One attractive feature of invariant automorphic forms, automorphic
functions, is that their structure and number closely reflect the
properties of the Lie algebra $\gg$. Once one takes the step to
transforming automorphic forms, the freedom is much bigger. Remember
that the scalar degrees of freedom reside in the coset
$G(\Z)\backslash G/K$, and that they appear through the ``physical''
part $P$ of the Maurer--Cartan form $g^{-1}dg$,
$P=\Pi_{\gg/\kk}(g^{-1}dg)$. Any higher-derivative term (considering
purely scalar terms) contain a number of $P$'s, perhaps with
covariant derivatives, contracted with something that cancels the
$K$ transformation of $P$ in the appropriate way. Note that
$\G(\m,g)$ transforms correctly, so that a $K$-invariant object may
be formed by contracting $P$'s either with each other, or with
$\G$'s. Again, summation over $G(\Z)$ is of course needed. We arrive
at automorphic forms of the generic form
$$
\phi^{(i,w,k)}_{\a_1\ldots\a_k}(g)=\sum_{\m\in G(\Z)}
\G_{\a_1}\ldots\G_{\a_k}C_i(\G)^{-w}\komma\eqn
$$
where again $\G_\a=\G_\a(\m,g)=[\Pi_{\gg/\kk}(g^{-1}\m g)]_\a$. The
automorphic form $\phi$ is symmetric in the $\a_k$ indices. The
restricted Casimir $C_i$ is inserted for convergence of the sum. We
see that, for a given choice of $C_i$ and $w$, there is one
automorphic form for each irreducible $K$-module in the symmetric
tensor product of $n$ elements in $\gg/\kk$, generically a much
larger number than the number of invariant automorphic forms.

Also here, the simpler construction for $G=SL(n)$ is obtained with
an even number of $(mg)_I$'s ($I$ is the fundamental index) as
$$
\psi^{(w,l)}_{I_1\ldots I_{2l}}(g)= \sum_{\Z^n\backslash
0}(mg)_{I_1}\ldots(mg)_{I_{2l}} \vert mg\vert^{-2w}\punkt\Eqn\SLaut
$$

We would like to comment on the transformation properties of the
transforming automorphic forms. As they are written (as functions of
$g$), they are a collection of functions on $G(\Z)\backslash G$,
transforming under $K$ transformations as specified by the index
structure. If we, on the other hand, view $g$ as a representative of
the right coset $G/K$ by fixing a gauge encoded in some
parametrisation $g=g(\tau)$, the picture changes. The coset
coordinates $\tau$ transform non-linearly under $G(\Z)$, and a
compensating gauge transformation is required to get back on the
gauge hypersurface. The transformations under $G(\Z)$ are of the
form $g(\tau)\rightarrow\g g(\tau)k(\g,\tau)$. In this picture, a
$G(\Z)$ transformation of the automorphic forms induces a
$K$-transformation with the element $k(\g,\tau)$ on the appropriate
module given by the index structure. This can of course be mimicked
without gauge fixing by replacing the element $\g\otimes1\in
G(\Z)\times K$ by the element $\g\otimes k(\g,\tau)$, which allows
us to interpret the automorphic forms as collections of functions on
$G/K$ with a specific non-linear transformation property under
$G(\Z)$.

We are sometimes interested in certain limiting values of
automorphic forms. In the present paper, the terms obtained after
dimensional reduction should correspond to leading terms in an
asymptotic expansion at large volume of a torus. We consider the
possibility of collecting the terms we obtain in sums of automorphic
functions of $SL(n+1)$ after reducing on $T^n$. With the embedding
of $GL(n)$ in $SL(n+1)$ discussed earlier in section \Rtwo,
where $e^\phi$ is the
determinant of the metric on $T^n$, the $SL(n+1)$ group element may
(with a partial gauge choice) be parametrised as
$$
G=\left[\matrix{e^{-\phi}&0\cr e^{-\phi}u&e^{\phi\over n}\tilde e\cr}\right]
\komma\eqn
$$
where $\tilde e$ is a group element of $SL(n)$ parametrising the
shape of $T^n$. The large-volume limit is $\phi\rightarrow\infty$.
The shape of $T^n$ should be irrelevant in this limit, as long
as it is non-degenerate, and we take
$\tilde e=\id$. An automorphic form of the type in eq. (\SLaut) with
$2w>2l+1$ (for convergence) is then dominated by terms with
$m=(m_0,0,\ldots,0)$ and and has the limiting value [\LambertWestII]
$$
\psi^{(w,l)}_{I_1\ldots I_{2l}}(g)
\mathop{\longrightarrow}_{\phi\rightarrow\infty}
e^{2(w-l)\phi}\zeta(2(w-l))\d_{I_1,0}\ldots\d_{I_{2l},0}\punkt\eqn
$$

Finally, it is interesting to count the number of possible terms one
can write down in a concrete situation. Much of the present paper
aims at reduction to $d=3$ and the coset $E_{8(8)}(\Z)\backslash
E_8/(Spin(16)/\Z_2)$. An $R^4$ correction contains terms with up to
eight $P$'s. Just considering these for a given $w$ (\ie, for the
moment omitting the terms with derivatives of $P$), and assuming
that we use the quadratic Casimir, the number of possible terms
obtainable are labeled by irreducible ${\frak so}(16)$ representations in
the symmetric product of eight chiral spinors. The number of
representations, \ie, of automorphic forms, is 222. This can be
compared to the number of scalars, 2, which is obtained directly
from the $E_8$ Casimir operators.
The corresponding number relevant for gravity, \ie, the number of
irreducible ${\frak so}(9)$ representations in the symmetric product of eight
symmetric traceless tensors, is 609. It seems that demanding
$E_{8(8)}$ invariance gives some restriction even on the possible
$SL(9,\Z)$-invariant terms involving the gravitational scalars only,
but it will take some ingenuity to extract the information.
It is tempting to believe that the
octic $E_8$ invariant [\CederwallPalmkvist] has some special r\^ole
in the $R^4$ terms, but this remains unclear in the light of the
large number of transforming automorphic functions.

\appendix{Reduction of the octic invariant to matrices}By
assuming that $E_{8}$ organises the scalars after compactification
to three dimensions also after the inclusion of $R^4$ terms we can
obtain constraints related to $SL(9)$ which are more readily
checked. To see this, consider the $\ee_8$ Dynkin diagram, with
Coxeter labels and extended root:

\vskip\parskip
\epsffile{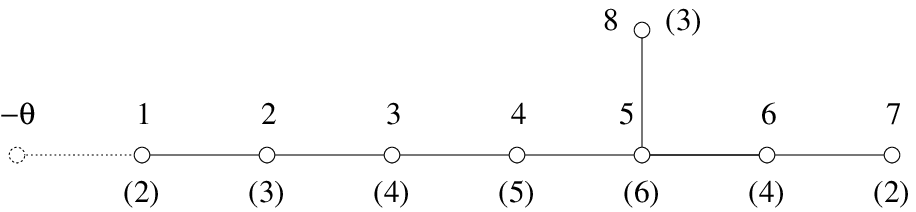}
\vskip\parskip

\noindent The horizontal line consists of the simple roots of ${\frak sl}(9)$.
In the standard way of embedding ${\frak sl}(n)$
roots in $(n+1)$-dimensional space,
an element in the Cartan algebra of ${\frak sl}(9)$
(and, thereby, of $\ee_8$) can
be written in an orthonormal basis as
$M=(m_0,m_1-m_0,m_2-m_1,\ldots,m_7-m_6,-m_7)$. We have
$\a_0=-\theta=-(2\a_1+3\a_2+4\a_3+5\a_4+6\a_5+4\a_6+2\a_7+3\a_8)$. Solving
for $\a_8$ gives
$\a_8=\fr3(-1,-1,-1,-1,-1,-1,2,2,2)$ in the orthonormal basis.

Invariants under ${\frak sl}(9)$ restricted to the CSA can be formed as
$\tr M^n\equiv\sum_{i=1}^9(M_i)^n$ (\ie, the vector $M$ above is thought of as
the diagonal of a matrix $M$). They will all be automatically invariant under
the Weyl group of ${\frak sl}(9)$, generated by simple reflections
permuting nearby components of the 9-dimensional vectors in the
orthonormal basis. The only thing one has to check for invariance under
the Weyl group of $\ee_8$ is invariance under reflection in the hyperplane
orthogonal to the exceptional root $\a_8$.
As a $(9\times9)$-matrix it is realised as
$$
w(\a_8)=\id-\a_8^t\a_8={1\over9}\left[\matrix{8&-1&-1&-1&-1&-1&2&2&2\cr
                          -1&8&-1&-1&-1&-1&2&2&2\cr
                          -1&-1&8&-1&-1&-1&2&2&2\cr
                          -1&-1&-1&8&-1&-1&2&2&2\cr
                          -1&-1&-1&-1&8&-1&2&2&2\cr
                          -1&-1&-1&-1&-1&8&2&2&2\cr
                          2&2&2&2&2&2&5&-4&-4\cr
                          2&2&2&2&2&2&-4&5&-4\cr
                          2&2&2&2&2&2&-4&-4&5\cr
}\right]\komma\Eqn\ExceptionalWeylElement
$$
and acts on $M$ as
$w(\a_8)M=M+\fr3m_5(-1,-1,-1,-1,-1,-1,2,2,2)$.
A general Ansatz for the restriction of the octic
$\ee_8$ invariant to the CSA (using
${\frak sl}(9)$ ``covariance'') is
$S(M)=\tr M^8+a\tr M^6\tr M^2+b\tr M^5\tr M^3+c(\tr M^4)^2
+d\tr M^4(\tr M^2)^2+e(\tr M^3)^2\tr M^2+f(\tr M^2)^4$.
The coefficient $f$ is of course arbitrary, and will be left out.
We demand that $S(w(\a_8)M)=S(M)$. A short Mathematica calculation then
gives the values of the coefficients in the Ansatz:
$$
\eqalign{
S(M)=\tr M^8&-\Fr{28}{45}\tr M^6\tr M^2-\Fr{28}{45}\tr M^5\tr M^3\cr
&-\Fr7{36}(\tr M^4)^2+\Fr7{36}\tr M^4(\tr M^2)^2+\Fr7{27}(\tr M^3)^2\tr M^2
\punkt\cr}\Eqn\SofM
$$
This is the polynomial (in the symmetric $(9\times9)$-matrix ${\scr P}$)
we should look for in the $R^4$ terms if it is multiplied
by a scalar automorphic form of $E_8$. It has to be the same formal
expression already in terms of the $P$ of $SO(8)$.

\acknowledgements We would like to thank Axel Kleinschmidt, 
Daniel Persson and Johan Bielecki for discussions
and Axel Kleinschmidt and Claudia Colonnello
for generously sharing unpublished results.

\vfill\eject
\refout

\end